\documentclass[twocolumn]{aastex631}

\usepackage[T1]{fontenc}
\usepackage{multirow}
\usepackage{hyperref}

\newcommand\teff{$T_{\mathrm{eff}}$}
\newcommand\logg{$\log g$}

\newcommand\gaia{\textit{Gaia}}
\newcommand\ot{O$_{\mathrm{2}}$}
\newcommand\ee{$\eta_{\oplus}$}
\newcommand\re{$R_{\oplus}$}

\shorttitle{Exo-Earth \ot\ with ELTs}
\shortauthors{Hardegree-Ullman et al.}

\graphicspath{{./}}

\begin{document}

\title{Bioverse: A Comprehensive Assessment of the Capabilities of Extremely Large Telescopes to Probe Earth-like \ot\ Levels in Nearby Transiting Habitable Zone Exoplanets}

\author[0000-0003-3702-0382]{Kevin K.\ Hardegree-Ullman}
\affiliation{Steward Observatory, The University of Arizona, Tucson, AZ 85721, USA; \href{mailto:kevinkhu@arizona.edu}{kevinkhu@arizona.edu}}

\author[0000-0003-3714-5855]{D\'{a}niel Apai}
\affiliation{Steward Observatory, The University of Arizona, Tucson, AZ 85721, USA; \href{mailto:kevinkhu@arizona.edu}{kevinkhu@arizona.edu}}
\affiliation{Lunar and Planetary Laboratory, The University of Arizona, Tucson, AZ 85721, USA}

\author[0000-0003-4500-8850]{Galen J.\ Bergsten}
\affil{Lunar and Planetary Laboratory, The University of Arizona, Tucson, AZ 85721, USA}

\author[0000-0001-7962-1683]{Ilaria Pascucci}
\affil{Lunar and Planetary Laboratory, The University of Arizona, Tucson, AZ 85721, USA}

\author[0000-0003-3204-8183]{Mercedes L\'opez-Morales} 
\affil{Center for Astrophysics ${\rm \mid}$ Harvard {\rm \&} Smithsonian, 60 Garden St, Cambridge, MA 02138, USA}

\begin{abstract}

Molecular oxygen is a strong indicator of life on Earth, and may indicate biological processes on exoplanets too. Recent studies proposed that Earth-like O$_2$ levels might be detectable on nearby exoplanets using high-resolution spectrographs on future extremely large telescopes (ELTs). However, these studies did not consider constraints like relative velocities, planet occurrence rates, and target observability. We expanded on past studies by creating a homogeneous catalog of 286,391 main-sequence stars within 120\,pc using \textit{Gaia} DR3, and used the \texttt{Bioverse} framework to simulate the likelihood of finding nearby transiting Earth analogs. We also simulated a survey of M dwarfs within 20\,pc accounting for $\eta_{\oplus}$ estimates, transit probabilities, relative velocities, and target observability to determine how long ELTs and theoretical 50--100 meter ground-based telescopes need to observe to probe for Earth-like O$_2$ levels with an $R=100,000$ spectrograph. This would only be possible within 50 years for up to $\sim$21\% of nearby M dwarf systems if a suitable transiting habitable zone Earth-analog was discovered, assuming signals from every observable partial transit from each ELT can be combined. If so, Earth-like O$_2$ levels could be detectable on TRAPPIST-1~d--g within 16 to 55 years, respectively, and about half that time with an $R=500,000$ spectrograph. These results have important implications for whether ELTs can survey nearby habitable zone Earth analogs for O$_2$ via transmission spectroscopy. Our work provides the most comprehensive assessment to date of the ground-based capabilities to search for life beyond the solar system.

\end{abstract}


\section{Introduction} \label{sec:intro}

Arguably, exoplanet science is driven by the search for life, or rather biosignatures, on planets beyond our solar system. On Earth, cyanobacteria, fueled by solar energy, started to generate molecular oxygen (\ot) as a waste product of oxygenic photosynthesis about 2.45 billion years ago. This oxygen accumulated in the ocean and eventually escaped to the atmosphere, displacing Earth's reducing atmosphere during the ``Great Oxidation Event'' \citep{Sessions2009}. In 1990, the \textit{Galileo} spacecraft obtained a spectrum of Earth which revealed \ot\ and methane (CH$_4$) in extreme thermodynamical disequilibrum, suggesting these two species are strong indicators of life on Earth \citep{Sagan1993}, and hence, potential biosignatures on other planets.

Since those observations, there have been in-depth investigations into the search for exoplanet biosignatures, in particular, the Nexus for Exoplanet System Science (NExSS) program organized a workshop in 2016 which yielded a set of exoplanet biosignature review articles summarizing the current state of the field \citep{Kiang2018,Meadows2018,Schwieterman2018,Catling2018,Fujii2018,Walker2018}. \citet{Meadows2018} focuses on \ot\ as a biosignature and how our understanding of \ot\ development and presence on Earth can lead to skewed interpretations of its detection on exoplanets. However, in spite of the challenges, \ot\ remains a primary biosignature, and there are several ways of ruling out false-positive scenarios, including the presence of CO and C\ot\ and the absence of CH$_4$ which would suggest an abiotic mechanism of \ot\ production \citep[e.g.,][]{Meadows2018,Krissansen-Totton2018,TheLUVOIRTeam2019}. We focus solely on \ot\ in this paper for two primary reasons: 1) other biosignatures such as CH$_4$, H$_2$O, and CO$_2$ should be detectable with facilities like JWST \citep[e.g.,][]{Wunderlich2019}, and 2) the ELTs are the only facilities currently under construction with a large enough aperture to search for \ot\ \citep{Lopez-Morales2019}.

Recently, \citet{Snellen2013}, \citet{Rodler2014}, \citet{Serindag2019}, and \citet{Lopez-Morales2019} studied the feasibility of using high-resolution spectrographs ($R\gtrsim100,000$) on upcoming extremely large telescopes (ELTs), such as the Giant Magellan Telescope (GMT), the Thirty Meter Telescope (TMT), and the European-Extremely Large Telescope (E-ELT) to detect oxygen in Earth analogs, i.e., Earth-like levels of \ot\ on habitable zone Earth-sized planets transiting nearby stars. \citet{Serindag2019} and \citet{Lopez-Morales2019} concluded that the best-case scenario to probe for \ot\ will be on an Earth-like planet transiting a mid-M dwarf (M4 or M5) at a distance between five and seven parsecs, which can yield a 3$\sigma$ Earth-like \ot\ measurement in 15--50 transits. Mid-M dwarfs constitute a majority of nearby stars and have the benefit of close-in habitable zones, higher geometric transit probability, short transit durations, and larger transit depths for small planets compared to FGK stars and early M dwarfs which could take hundreds of transits to get a significant \ot\ signal. Habitable zone terrestrial planets orbiting FGK stars are much better suited to direct imaging observations from future large-aperture space-based telescopes for atmospheric characterization \citep[e.g.,][]{TheLUVOIRTeam2019,Gaudi2020}. In this paper, we further investigate the feasibility and likelihood of measuring Earth-like levels of \ot\ in the atmospheres of nearby transiting exoplanets in the near future with ELTs.

In order to provide a comprehensive assessment of potential exoplanet targets for the observations studied, we developed and added a new stellar catalog to the \texttt{Bioverse}\footnote{\url{https://github.com/abixel/bioverse}} framework \citep{Bixel2021}. Previously, the \texttt{Bioverse} stellar catalog was generated randomly from a present-day stellar mass function \citep{Chabrier2003} because the census of low-mass stars was not complete out to $\sim$100\,pc, but that problem has now been mitigated with \gaia. Including our new catalog in the open-source \texttt{Bioverse} enables future studies of broadly similar science questions to those presented here and provides a real stellar population instead of a hypothetical population of nearby stars based on models. 

In Section~\ref{sec:starpars}, we review the steps taken to build our stellar catalog. In Section~\ref{sec:o2}, we explore the nearby stellar population, their potential for hosting Earth-sized habitable zone planets, and their observability from ground-based telescopes. Next, we use models of expected ELT high-resolution spectrograph performance and capabilities to compute how long it will take to probe for Earth-like levels of \ot\ if we find an exoplanet suitable for such measurements in Section~\ref{sec:time}. Finally, we summarize our results and conclude with recommendations in Section~\ref{sec:summary}.

\section{Stellar Parameters} \label{sec:starpars}

Fundamental stellar parameters, including effective temperature (\teff), luminosity ($L_{\star}$), radius ($R_{\star}$), and mass ($M_{\star}$) are essential ingredients for basic analysis of exoplanet systems. From \teff\ and $L_{\star}$ we can measure stellar incident flux and equilibrium temperature to determine an exoplanet's potential for habitability. Stellar radius and stellar mass are necessary to measure planet radius and mass, which are critical to learn about the atmosphere and composition of an exoplanet. There are many different methods to measure these fundamental properties (e.g., spectroscopy, interferometry, modeling, etc.), however, in the absence of precise measurements for every star, we can estimate properties for thousands of stars from photometry and astrometry.

The base catalog for this study is the \gaia\ Catalogue of Nearby Stars, which we introduce in Section~\ref{sec:gcns}. We began our analysis by computing photometric colors and absolute magnitudes and removing non-main sequence stars (Section~\ref{sec:colmag}). We then used the colors and absolute magnitudes to measure \teff\  (Section~\ref{sec:teff}). Next, we used \teff\ and absolute magnitudes to compute $L_{\star}$ (Section~\ref{sec:lum}), and with these parameters at hand, used the Stefan-Boltzmann law to determine $R_{\star}$ (Section~\ref{sec:rad}). Finally, we used mass-luminosity relationships to measure $M_{\star}$ (Section~\ref{sec:mass}).

In addition to basic stellar parameters, more in-depth analysis of a star system may be necessary. For example, it is helpful to know if an observed star has a close-by stellar companion, which can dilute any measurements of planets found in the system \citep[e.g.,][]{Furlan2017}. We explore stellar multiplicity in Section~\ref{sec:binary}. Also relevant to ground-based biosignature searches is the relative velocity of a star system with respect to Earth, which requires stellar systemic velocity measurements (Section~\ref{sec:sv}). We added all of our measured and compiled stellar measurements into the \texttt{Bioverse} framework and used them to explore the possibility of measuring \ot\ in the atmosphere of an Earth-like planet within 20\,pc using upcoming ELTs (Section~\ref{sec:o2}).
\newline
\subsection{Gaia Catalogue of Nearby Stars} \label{sec:gcns}

The \gaia\ Catalogue of Nearby Stars (GCNS) contains 331,312 targets out to 120\,pc from \gaia\ EDR3 \citep{GaiaCollaboration2016,GaiaCollaboration2021}, and is estimated to be nearly 100\% complete out to 100\,pc from the Sun through spectral type M7 and 100\% complete out to 50\,pc through spectral type M9 \citep{GCNS2021}. While \citet{GCNS2021} found the GCNS to be nearly complete, they identified 1,259 objects on SIMBAD\footnote{\url{http://SIMBAD.u-strasbg.fr}} \citep{Wenger2000} with a parallax larger than 10 milliarcseconds that are not in the GCNS. These notably include some of our closest neighboring stars, such as $\mathrm{\alpha}$ Centauri A and B. As \citet{GCNS2021} notes, many of these stars are extremely faint, extremely bright, or binaries. Some, however, have no clearly identified reason to be missing from GCNS. For completeness, we include these 1,259 targets in our catalog.

The GCNS contains distance measurements and compiled optical ($G$, $G_{BP}$, $G_{RP}$) photometry. The GCNS was compiled using \gaia\ EDR3, but the photometric passband definitions between EDR3 and DR3 remained the same.\footnote{\url{https://www.cosmos.esa.int/web/gaia/dr3}} However, $G$-band photometry for faint ($G < 13$) \gaia\ EDR3 sources required a minor correction \citep{Riello2021,GaiaCollaboration2021}. We used the corrected $G$-band photometry in our work. While the GCNS also contains additional compiled optical ($g$, $r$, $i$, $z$) and infrared ($J$, $H$, $K_S$, $W1$, $W2$, $W3$, $W4$) photometry, we primarily used the \gaia-band photometry since it is the most complete. Over 90\% of GCNS stars have $K_S$-band magnitude measurements, which we used in our characterization of M dwarfs (see Sections~\ref{sec:rad} and \ref{sec:mass}). For completeness, we queried SIMBAD for parallaxes and photometry for the aforementioned 1,259 targets not in the GCNS. About 58\% of these targets have \gaia-band photometry from either \gaia\ DR3 or \gaia\ DR2, 70\% have $V$-band photometry, and 89\% have $K_S$-band photometry. Due to the nature of the \gaia\ mission, passband definitions between data releases are not identical.\footnote{\url{https://www.cosmos.esa.int/web/gaia/edr3-passbands}} For some of our targets and analysis, we cautiously used information from \gaia\ DR2 due to lack of new or updated data from \gaia\ DR3, and because the typical difference between DR2 and DR3 photometry for our targets is very small (0.02 magnitudes).
\newline

\subsection{Colors and Absolute Magnitudes} \label{sec:colmag}

\begin{figure*}[ht!]
\includegraphics[width=0.99\linewidth]{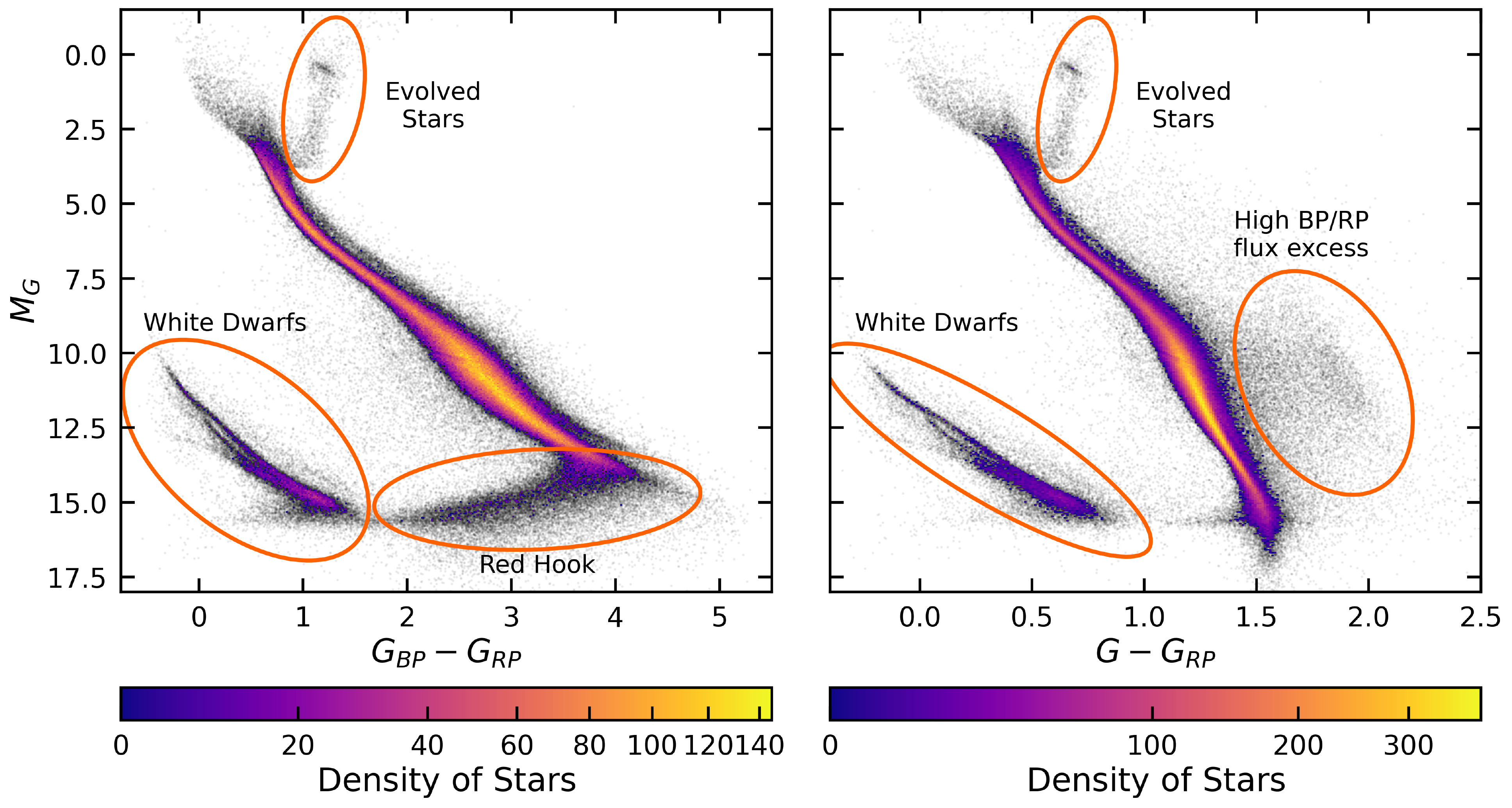}
\includegraphics[width=0.99\linewidth]{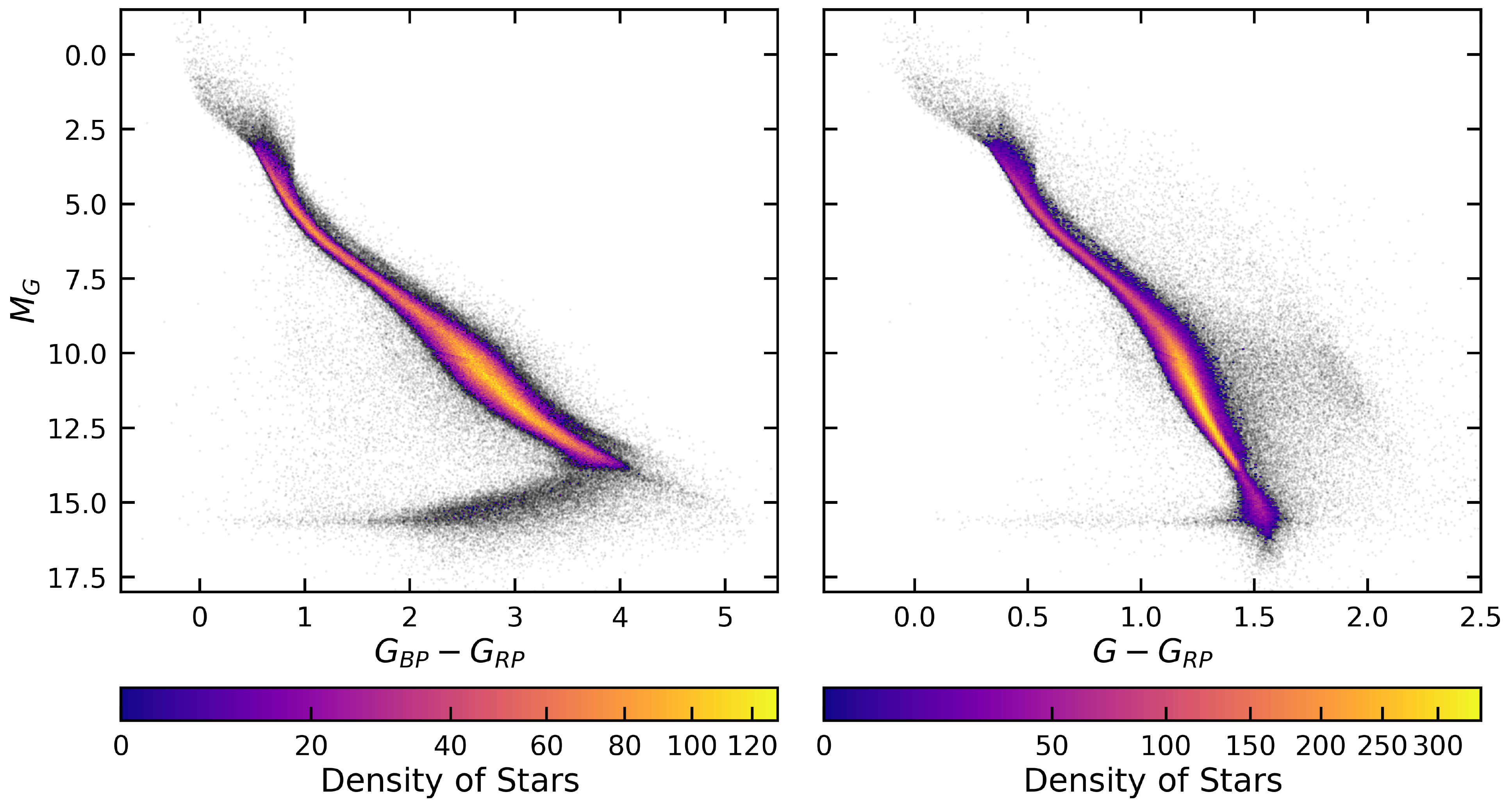}
\caption{Color-magnitude diagrams of GCNS targets. The upper two panels show all GCNS targets with $G$, $G_{BP}$ and $G_{RP}$ measurements. Red ellipses indicate the locations of evolved stars and white dwarfs in both plots, the red hook in the $G_{BP}-G_{RP}$ plot, and a region of high BP/RP flux excess in the $G-G_{RP}$ plot. The lower two panels show only main-sequence stars used in this study. \label{fig:gcnscmd}}
\end{figure*}

The fist step was to compute absolute $G$, $G_{BP}$, $G_{RP}$, $V$, and $K_S$ magnitudes and $G_{BP}-G_{RP}$ and $G-G_{RP}$ colors. Since our targets are nearby, interstellar reddening is not expected to be significant \citep[$\approx$0.01~mag at 100\,pc;][]{Aumer2009}, so we do not include it in our calculations of absolute magnitudes. Figure~\ref{fig:gcnscmd} shows color-magnitude diagrams of these targets. At magnitudes fainter than $M_G\approx13$, there is a notable hook feature in the $G_{BP}-G_{RP}$ versus $M_G$ plot where the main sequence appears to turn toward the white dwarfs. This feature is an artifact of poorly estimated $G_{BP}$ measurements for intrinsically red targets. This feature is muted in the $G-G_{RP}$ versus $M_G$ plot, but there is a cloud of points to the right of the main sequence that have high BP/RP flux excess values, which is not present in the other plot because $G_{BP}$ and $G_{RP}$ are biased by the same amount \citep[see also Figure 29 of][]{GCNS2021}. Since erroneous color measurements would lead to inaccurate stellar parameter derivations, we applied a magnitude limit to choose the color--\teff\ relationship we used, which is explained in detail in Section~\ref{sec:teff}. This allowed us to create a more complete nearby stellar catalog by including stars through the end of the main sequence. Before we could use our computed colors and absolute magnitudes to compute stellar parameters, we performed some cuts to remove non-main sequence stars as explained in the following section.

\subsection{Removing Non-Main Sequence Stars} \label{sec:nonms}

Since our stellar classifications are tuned to main-sequence stars, any measurements for other stars will not be accurate, so we removed them from our target list. First, \citet{GCNS2021} computed a white dwarf probability for catalog targets using known white dwarfs and non-white dwarfs, \gaia\ photometry, proper motions, parallaxes, and parallax uncertainties as a training set for a random forest classification algorithm. This classifier algorithm correctly identified over 98\% of white dwarfs and over 99\% of the non-white dwarfs from the training set. This classifier was used on the entire GCNS catalog to assign a white dwarf probability \citep{GCNS2021}. We used this metric to remove targets with a white dwarf probability higher than 50\%, as was done in \citet{GCNS2021} when they derived a local luminosity function. Second, we removed evolved stars with $G_{BP}-G_{RP}>0.9$ and $M_G<4$ (Figure~\ref{fig:gcnscmd} top left panel). Lastly, we identified and removed additional giants, white dwarfs, brown dwarfs, and young stars using object types obtained from SIMBAD. The lower two panels of Figure~\ref{fig:gcnscmd} show color-magnitude diagrams of the stellar population we use in this study after removal of non-main sequence stars.

\subsection{Effective Temperature} \label{sec:teff}

One of the most fundamental parameters for determining the location of the habitable zone around main-sequence stars is the stellar temperature. To uniformly determine the effective temperatures of the stars in our target catalog, we fit polynomials to the colors and \teff\ or absolute magnitudes and \teff\ from Table~5 of \citet{Pecaut2013}.\footnote{We used the most up-to-date information from \url{http://www.pas.rochester.edu/\~emamajek/EEM_dwarf_UBVIJHK_colors_Teff.txt} (as of April 2023), extracting individual measurements of \teff\ for each spectral type from \url{http://www.pas.rochester.edu/~emamajek/spt/} in order to aid in computing uncertainties.} The order of each polynomial in this work was selected using the Bayesian Information Criterion \citep[BIC,][]{Schwarz1978}, iterating polynomial fits between order two and 15 and selecting the order with the lowest BIC. We calculated \teff\ for each target using each color and magnitude listed above in Section~\ref{sec:colmag}. We computed uncertainty from the following two components added in quadrature. First, we used a Monte Carlo method to propagate magnitude measurement uncertainties through each calculation, drawing 100 magnitude values from a Gaussian distribution, fitting the polynomial to each color or magnitude distribution, and selecting the median as the \teff\ and standard deviation as the first component in the uncertainty. Second, we computed the standard deviation of the residuals of the polynomial fit divided by the median \teff, which was then multiplied by the computed \teff\ as the second component in the uncertainty. In order of preference, we selected a final \teff\ from (1) $G_{BP}-G_{RP}$ for targets with $M_G\le12.5$, (2) $G-G_{RP}$ for targets with $M_G>12.5$, (3) $M_G$, (4) $M_{K_S}$, and (5) $M_V$. For brighter targets, using $G_{BP}-G_{RP}$ avoided issues of high BP/RP flux excess as seen in $G-G_{RP}$, and for fainter targets using $G-G_{RP}$ minimized the red hook effect as seen in $G_{BP}-G_{RP}$. Whenever possible we used the color-temperatures over the magnitude-temperatures due to a much lower order polynomial fits (fourth or fifth order versus tenth to twelfth order), but colors were not available for all targets. The median uncertainty on \teff\ is 3.3\%, which is above the realistic systematic uncertainty floor of $\approx$2.4\% found by \citet{Tayar2022} based on measurements from interferometric angular diameters and bolometric fluxes for FGK main-sequence stars.

\subsection{Luminosity} \label{sec:lum}

In order to compute stellar luminosities, we first computed bolometric corrections (BC). We used the \gaia\ DR3 bolometric correction tool\footnote{\url{https://gitlab.oca.eu/ordenovic/gaiadr3_bcg}}, which is described in \citet{Creevey2022}. This tool computes BC$_G$ from synthetic stellar spectra from inputs of \teff, \logg, [Fe/H], and [$\alpha$/Fe]. Since we do not have a large, uniform catalog of our own measurements for \logg, [Fe/H], and [$\alpha$/Fe], we had to make the following assumptions to minimize our inputs to the BC$_G$ tool.

We first reduced our reliance on \logg\ by establishing a \teff-\logg\ relationship using Table~5 of \citet{Pecaut2013}. For stars between 2,500 and 10,000\,K (corresponding to the temperature range of AFGKM stars used in this study), we computed $\log g=\log(GM_{\star}/R_{\star}^2)$, where $G$ is the gravitational constant and $M_{\star}$ and $R_{\star}$ are the stellar masses and radii from \citet{Pecaut2013}. %for our temperature range.
We computed \logg\ by fitting a third degree polynomial to \teff\ versus \logg, and used that computed \logg\ in the BC$_G$ tool.

Next, we needed to make an assumption regarding stellar metallicity to compute BC$_G$. Metallicities of large populations of stars in the solar neighborhood (within 120\,pc) have been measured for both FGK stars \citep[e.g.,][]{Haywood2001,Casagrande2011,Sousa2011} and M dwarfs \citep[e.g.][]{Terrien2015,Dittmann2016}. These studies yielded slightly sub-solar [Fe/H] for nearby FGK stars ($-0.04^{+0.19}_{-0.25}$, $-0.07\pm0.19$, and $-0.08^{+0.20}_{-0.24}$ from \citet{Haywood2001}, \citet{Casagrande2011}, and \citet{Sousa2011}, respectively) and slightly super-solar [Fe/H] for nearby M dwarfs ($0.06^{+0.25}_{-0.21}$ and $0.02^{+0.18}_{-0.27}$ from \citet{Terrien2015} and \citet{Dittmann2016}, respectively). Because of the relatively large spread in the [Fe/H] measurements from each study, they all yield similar results of effectively solar metallicity.

We ran a Monte Carlo simulation, randomly selecting [Fe/H] and [$\alpha$/Fe] (from the same star) from \citet{Casagrande2011} 100 times for each effective temperature between 2,500 and 10,000\,K at intervals of 50\,K, running these values through the $Gaia$ DR3 bolometric correction tool along with the \logg\ at each associated \teff\ from the previous step. We selected \citet{Casagrande2011} because it was the only catalog with both [Fe/H] and [$\alpha$/Fe], and contained an order-of-magnitude more measurements than each of the other catalogs. We then fit a seventh order polynomial to the resultant \teff\ and BC$_G$ values from our simulation, which yielded a root mean square error (RMSE) of 0.03 magnitudes. We adopted this value as the uncertainty on the measurement of BC$_G$ for use in our calculation of $L_{\star}$ below.

These bolometric corrections are for the \gaia\ $G$-band, but not all targets have $G$-band measurements. For the remaining targets with $V$-band measurements, we fit a tenth order polynomial to the \teff\ and BC$_V$ values from \citet{Pecaut2013}, which again yielded a $\sim$0.03 magnitude RMSE uncertainty for BC$_V$. With bolometric corrections in hand, luminosity ($L_{\star}$) was computed using
\begin{equation}
    L_{\star} = 10^{-0.4(M_{G,V}+\mathrm{BC}_{G,V}-M_{\mathrm{bol},\odot})}
\end{equation}
where $M_{\mathrm{bol},\odot} = 4.74$ mag is the solar bolometric magnitude \citep{Mamajek2015}. Uncertainties were again calculated using the aforementioned Monte Carlo method, yielding a median luminosity uncertainty of 6\% \citep[compared to the systematic uncertainty floor for FGK stars of $\approx$2\% from][]{Tayar2022}.

\subsection{Radius} \label{sec:rad}

Since we now have \teff\ and luminosity, stellar radius is computed from the Stefan-Boltzmann law
\begin{equation}
    \frac{R_{\star}}{R_{\odot}} = \left(\frac{T_{\mathrm{eff}}}{T_{\odot}}\right)^{-2}\left(\frac{L_{\star}}{L_{\odot}}\right)^{1/2}
\end{equation}
where $R_{\odot}$, $T_{\odot}$, and $L_{\odot}$ are the solar values for radius, temperature, and luminosity. The median uncertainty on $R_{\star}$ is 6.8\% \citep[compared to the systematic uncertainty floor for FGK stars of $\approx$4.4\% from][]{Tayar2022}. For targets with $4.5 < M_{K_S} < 10$, we preferred to use the $M_{K_S}$--$R_{\star}$ relations from \citet{Mann2015}, which are finely tuned for M dwarfs and yield radius uncertainties around 3\%.

\subsection{Mass} \label{sec:mass}

We used the mass-luminosity relation calibrated from 95 detached binary systems from \citet{Torres2010} for stars with $\log(L_{\star}/L_{\odot})>-1$, which corresponds to a mass $M_{\star}\gtrsim0.7\,M_{\odot}$. The median uncertainty on $M_{\star}$ is 7.9\% \citep[compared to the systematic uncertainty floor for FGK stars of $\approx$5\% from][]{Tayar2022}. We then used an $M_{K_S}$--$M_{\star}$ relation for M dwarfs with $4.5 < M_{K_S} < 10$ from \citet{Mann2019}, which yields 3.2\% uncertainties.

Since not all stars fall within the above two categories, we used a random forest regression algorithm \citep{Breiman2001,Pedregosa2011} to predict $M_{\star}$ for the remaining 16,138 stars. Random forest regression for stellar classification has been used in a similar manner by \citet{Hardegree-Ullman2019} and \citet{Hardegree-Ullman2020}. We used targets with \teff, $L_{\star}$, $R_{\star}$, and $M_{\star}$ measurements in order to predict $M_{\star}$ for targets with only \teff, $L_{\star}$ and $R_{\star}$ measurements. We randomly selected 75\% of our input targets as a training set for the random forest algorithm. The remaining 25\% of the sample was used to test the effectiveness of the random forest algorithm. The RMSE between measured and predicted $M_{\star}$ values was only 0.3\%, which we added in quadrature to the 7.9\% median uncertainty for FGK stars from above.

In total, we are left with 286,391 stars with complete \teff, $L_{\star}$, $R_{\star}$, and $M_{\star}$ measurements. These derived parameters are provided in Table~\ref{tab:starpars}.

\subsection{Stellar Companions} \label{sec:binary}

Between 50 and 75\% of Sun-like FGK stars and about 30\% of M dwarfs within 25\,pc are in multiple star systems \citep{Raghavan2010,Winters2019}. Unresolved stellar companions pose a problem for interpreting exoplanet signals since inaccurate stellar parameters assuming single stars will often affect the derived planet parameters \citep[e.g.,][]{Kraus2016,Furlan2017}. In order to identify known and probable multiple star systems, we used the resolved GCNS multiple systems catalog \citep{GCNS2021}, the \gaia\ EDR3 binary catalog \citep{El-Badry2021}, the Robo-AO census of companions within 25\,pc \citep{Salama2022}, the \gaia\ DR3 non-single stars catalog \citep{GaiaCollaboration2022}, and any targets with the object type identifier of eclipsing binary or spectroscopic binary from SIMBAD. If a target was identified to be in a multiple star system from the above catalogs, we added a flag to Table~\ref{tab:starpars}, and we advise caution using any of our derived stellar parameters for these targets. We did not include binary systems in our \ot\ analysis below due to their potential to suppress planet formation at close separations \citep[e.g.,][]{Wang2014}, and the above empirical relationships for stellar properties were derived from single stars or detached binaries and assumed calculations for single stars, which would be wrong if there is flux contamination from a nearby companion. Binary systems remain in the catalog in case someone wants to work them and can correct for effects such as flux dilution \citep[e.g.,][]{Ciardi2015}. We found a stellar companion rate of 30\% for FGK dwarfs and 14\% for M dwarfs, which is about half of the rate of the literature values from \citet{Raghavan2010} and \citet{Winters2019}. Since \citet{Raghavan2010} and \citet{Winters2019} were volume-limited surveys (25\,pc) and our sample is out to 120\,pc, it is likely our compiled list of binaries is incomplete, especially at further distances.

\subsection{Systemic Velocities} \label{sec:sv}

Because the ground-based searches for \ot\ absorption on exoplanets (through the terrestrial atmosphere) rely on the Doppler effect to shift the exoplanet's \ot\ absorption lines out of the terrestrial ones, the relative velocities of Earth and the targeted systems are an important consideration for detectability \citep{Rodler2014,Lopez-Morales2019}. To uniformly assess the relative velocities, we used
\gaia\ DR3, which contains spectroscopic systemic velocities of nearly 34 million stars with temperatures between 3,100 and 14,500 K \citep[][]{Katz2022}. These measurements are available for 60\% of our stars. Systemic velocity measurements from the literature are available on SIMBAD for an additional 1.2\% of our targets. Astrometric systemic velocity measurements can be made by monitoring changing parallax or changing proper motions, but uncertainties tend to be relatively large \citep[e.g.,][]{Dravins1999,Lindegren2021}.

\begin{figure}[ht!]
\includegraphics[width=0.99\linewidth, trim=2 2 2 2, clip]{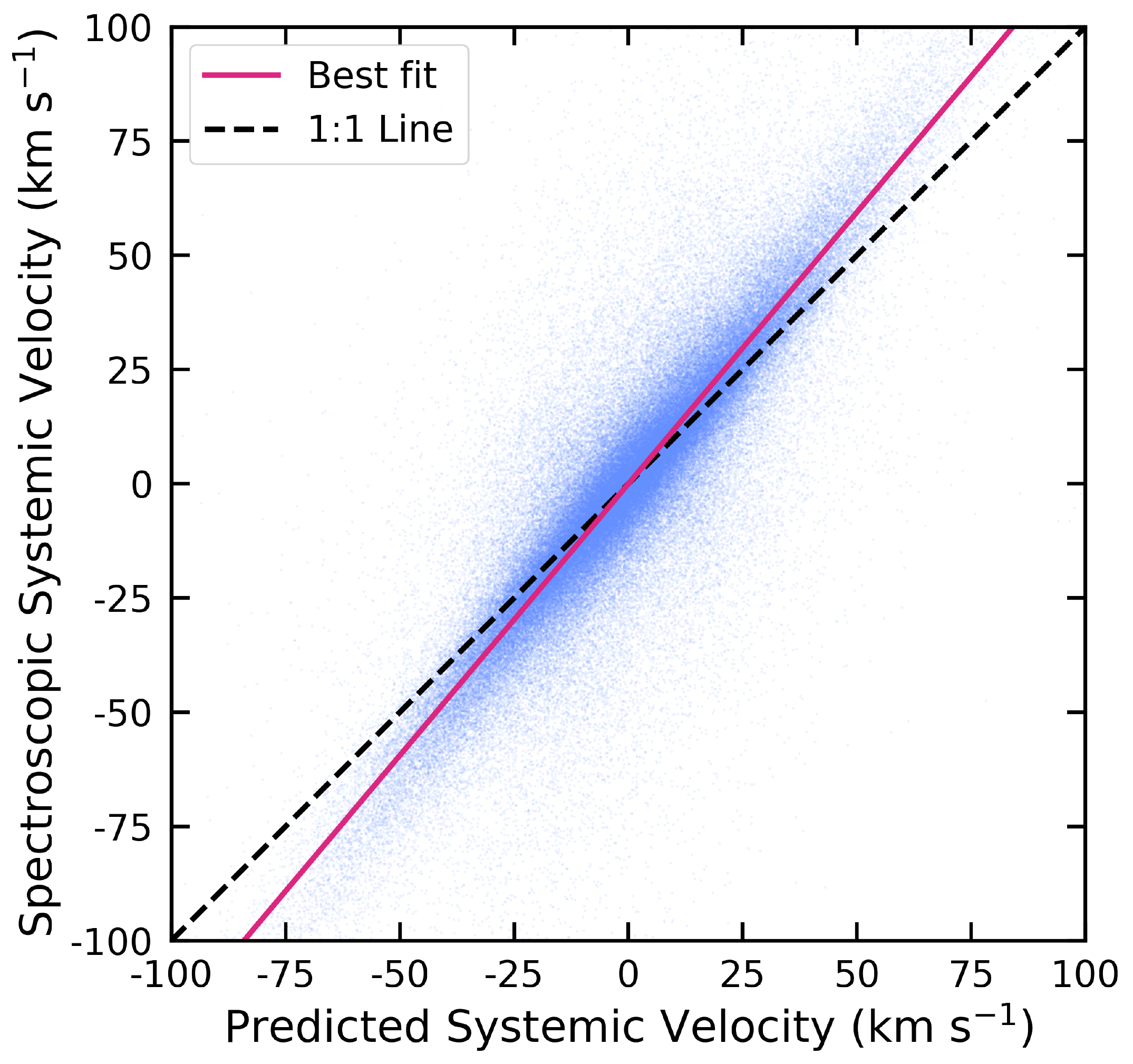}
\caption{Comparison between spectroscopic and predicted systemic velocity measurements from random forest regression. A 1:1 correlation is shown as a dashed black line, and the line of best fit through the data is shown as a solid pink line. We used the line of best fit to apply an offset to all predicted systemic velocity measurements without spectroscopic measurements.\label{fig:sv}}
\end{figure}

We applied a novel approach to determining astrometric systemic velocities, similar to our methodology in Section~\ref{sec:mass}, we used a random forest regression algorithm trained on targets with spectroscopic systemic velocities. We used the right ascension ($\alpha$), declination ($\delta$), proper motions ($\mu_{\alpha}$, $\mu_{\delta}$), and stellar distance for targets with spectroscopic systemic velocities to predict astrometric systemic velocities. Figure~\ref{fig:sv} shows the predicted astrometric systemic velocities compared to the spectroscopic systemic velocities. There is a strong positive correlation, slightly offset from a one-to-one correlation due to the smaller number of of targets available at high absolute systemic velocities for the training set. We applied the trained random forest algorithm to targets without spectroscopic systemic velocities and offset the measurements using the line of best fit shown in Figure~\ref{fig:sv}. For these targets, we adopted a one sigma uncertainty of 17.51 km s$^{-1}$ based on the RMSE from the residual of the fit to the spectroscopic versus predicted systemic velocity measurements. Now, 99.95\% of our targets have systemic velocity measurements. The remaining 0.05\% of targets did not have the proper motion measurements necessary to make a systemic velocity prediction. Our derived stellar parameters can be found in Table~\ref{tab:starpars} along with a column indicating if the systemic velocity measurement is from \gaia, SIMBAD, or our predicted values.

We have incorporated our stellar catalog into \texttt{Bioverse} as the new base stellar catalog for calculations of exoplanet systems. Since our new stellar catalog is based on the GCNS, we are now mostly complete out to $\sim$100\,pc. We left the ability to generate a stellar catalog based on the stellar mass function from \citep{Chabrier2003} in \texttt{Bioverse} for simulations and in order to exactly reproduce the results of \citet{Bixel2021}.

\begin{deluxetable*}{hhlhhhhhhhhhclccchhhhhhcrc}
\tabletypesize{\scriptsize}

\tablecaption{Photo-astrometric properties of main sequence stars within 120\,pc.\label{tab:starpars}}

\tablehead{\nocolhead{Gaia DR3} & \nocolhead{2MASS} & \colhead{Name} & \nocolhead{RA (J2000)} & \nocolhead{Dec (J2000)} & \nocolhead{PM RA} & \nocolhead{PM Dec} & \nocolhead{$G$} & \nocolhead{$BP$} & \nocolhead{$RP$} & \nocolhead{$V$} & \nocolhead{$K_s$} & \colhead{Distance} & \colhead{$T_{\mathrm{eff}}$} & \colhead{$L_{\star}$} & \colhead{$R_{\star}$} & \colhead{$M_{\star}$} & \nocolhead{RUWE} & \nocolhead{GCNS binary} & \nocolhead{Gaia EDR3 binary} & \nocolhead{Robo-AO binary} & \nocolhead{Gaia DR3 binary} & \nocolhead{SIMBAD binary} & \colhead{Binary} & \colhead{Systemic Velocity} & \colhead{Systemic Velocity}\vspace{-8pt} \\
\nocolhead{} & \nocolhead{} & \colhead{} & \nocolhead{($^{\circ}$)} & \nocolhead{($^{\circ}$)} & \nocolhead{(mas/yr)} & \nocolhead{(mas/yr)} & \nocolhead{(mag)} & \nocolhead{(mag)} & \nocolhead{(mag)} & \nocolhead{(mag)} & \nocolhead{(mag)} & \colhead{(pc)} & \colhead{(K)} & \colhead{($L_{\odot}$)} & \colhead{($R_{\odot}$)} & \colhead{($M_{\odot}$)} & \nocolhead{} & \nocolhead{} & \nocolhead{} & \nocolhead{} & \nocolhead{} & \nocolhead{} & \colhead{} & \colhead{(km s$^{-1}$)} & \colhead{Reference$^{\dagger}$} } 

\startdata
5853498713190525696 & 14294291-6240465 & Proxima Centauri & 217.4289422 & -62.6794902 & $-3781.7410\pm0.0314$ & $769.4650\pm0.0505$ & $8.9847\pm0.0028$ & $11.3731\pm0.0038$ & $7.5685\pm0.0042$ & $11.130\pm0.990$ & $4.384\pm0.033$ & $1.300^{+0.000}_{-0.020}$ & $2896\pm79$ & $0.00136\pm0.00004$ & $0.147\pm0.008$ & $0.117\pm0.009$ & 0.97 & F & F & F & F & F & F & $-21.9427\pm0.2161$ & 1 \\
 & 14393592-6050069 & * alf Cen A & 219.9020583 & -60.8339927 & $-3679.2500\pm3.2400$ & $473.6700\pm3.8900$ &  &  &  & $0.010\pm0.990$ & $-2.008\pm0.260$ & $1.346^{+0.003}_{-0.002}$ & $6453\pm390$ & $1.47279\pm1.47279$ & $1.005\pm0.331$ & $1.129\pm0.294$ &  & F & F & F & F & T & T & $-21.4000\pm0.7600$ & 2 \\
 &  & * alf Cen B & 219.8960963 & -60.8375276 & $-3614.3900\pm19.5200$ & $802.9800\pm20.4800$ &  &  &  & $1.330\pm0.990$ & $-0.600\pm0.990$ & $1.346^{+0.002}_{-0.002}$ & $5042\pm980$ & $0.46149\pm0.46149$ & $0.971\pm0.497$ & $0.795\pm0.206$ &  & F & F & F & F & F & F & $-22.5860\pm0.0001$ & 2 \\
4472832130942575872 & 17574849+0441405 & Barnard's star & 269.4520769 & 4.6933650 & $-801.5510\pm0.0318$ & $10362.3942\pm0.0361$ & $8.1940\pm0.0028$ & $9.7918\pm0.0030$ & $6.9581\pm0.0038$ & $9.511\pm0.990$ & $4.524\pm0.020$ & $1.840^{+0.010}_{-0.010}$ & $3263\pm109$ & $0.00344\pm0.00011$ & $0.181\pm0.013$ & $0.148\pm0.012$ & 1.08 & F & F & F & F & F & F & $-110.4682\pm0.1313$ & 1 \\
3864972938605115520 & 10562886+0700527 & Wolf  359 & 164.1205036 & 7.0147231 & $-3866.3383\pm0.0813$ & $-2699.2150\pm0.0691$ & $11.0384\pm0.0031$ & $13.7703\pm0.0054$ & $9.5854\pm0.0043$ & $13.507\pm0.006$ & $6.084\pm0.017$ & $2.410^{+0.000}_{-0.040}$ & $2823\pm77$ & $0.00081\pm0.00003$ & $0.135\pm0.004$ & $0.108\pm0.003$ & 0.84 & F & F & F & F & F & F & $19.3210\pm0.1450$ & 1 \\
762815470562110464 & 11032027+3558203 & HD  95735 & 165.8341451 & 35.9698822 & $-580.0571\pm0.0256$ & $-4776.5887\pm0.0300$ & $6.5512\pm0.0028$ & $7.6911\pm0.0030$ & $5.4755\pm0.0039$ & $7.520\pm0.009$ & $3.501\pm0.352$ & $2.550^{+0.000}_{-0.050}$ & $3568\pm119$ & $0.02327\pm0.00087$ & $0.389\pm0.048$ & $0.368\pm0.056$ & 0.96 & F & F & F & F & F & F & $-85.1106\pm0.1310$ & 1 \\
5140693571158946048 &  & G 272-61B & 24.7568238 & -17.9502782 & $3178.6943\pm0.4269$ & $584.0613\pm0.3023$ & $10.8178\pm0.0034$ & $13.2522\pm0.0104$ & $9.4237\pm0.0158$ &  &  & $2.670^{+0.010}_{-0.000}$ & $2938\pm84$ & $0.00097\pm0.00003$ & $0.118\pm0.008$ & $0.093\pm0.007$ & 10.45 & F & F & F & F & F & F & $10.2495\pm3.2382$ & 1 \\
5140693571158739840 &  & G 272-61A & 24.7557386 & -17.9507188 & $3385.3158\pm0.6722$ & $544.3864\pm0.3788$ & $10.5075\pm0.0029$ & $12.8206\pm0.0062$ & $9.0677\pm0.0130$ &  &  & $2.720^{+0.010}_{-0.010}$ & $2845\pm80$ & $0.00153\pm0.00005$ & $0.161\pm0.010$ & $0.130\pm0.010$ & 12.37 & F & F & F & F & F & F & $21.4582\pm0.7901$ & 1 \\
4075141768785646848 & 18494929-2350101 & Ross  154 & 282.4556824 & -23.8362353 & $639.3679\pm0.0368$ & $-193.9579\pm0.0318$ & $9.1264\pm0.0029$ & $10.7322\pm0.0042$ & $7.8981\pm0.0040$ & $10.430\pm0.010$ & $5.370\pm0.016$ & $2.980^{+0.030}_{-0.000}$ & $3263\pm109$ & $0.00389\pm0.00013$ & $0.211\pm0.006$ & $0.178\pm0.006$ & 0.99 & F & F & F & F & F & F & $-11.1164\pm0.5655$ & 1 \\
1926461164913660160 &  & Ross  248 & 355.4793180 & 44.1774497 & $112.5275\pm0.0364$ & $-1591.6498\pm0.0268$ & $10.3793\pm0.0028$ & $12.5464\pm0.0035$ & $9.0170\pm0.0038$ &  &  & $3.160^{+0.060}_{-0.000}$ & $2994\pm81$ & $0.00191\pm0.00006$ & $0.161\pm0.009$ & $0.130\pm0.010$ & 1.03 & F & F & F & F & F & F & $-77.2898\pm0.1853$ & 1 \\
\enddata

\tablerefs{$\dagger$ (1) Gaia DR3 \citep{Katz2022}, (2) Literature values from SIMBAD \citep{Wenger2000}, (3) This work (Section~\ref{sec:sv}).}

\tablecomments{Table~\ref{tab:starpars} is published in its entirety in machine-readable format with additional columns. A small portion is shown here for guidance regarding its form and content.}
\vspace{-24pt}
\end{deluxetable*}
\vspace{-28pt}

\subsection{Stellar Parameter Comparison to Previous Measurements}\label{sec:starparcomp}

\begin{figure*}[ht!]
\plottwo{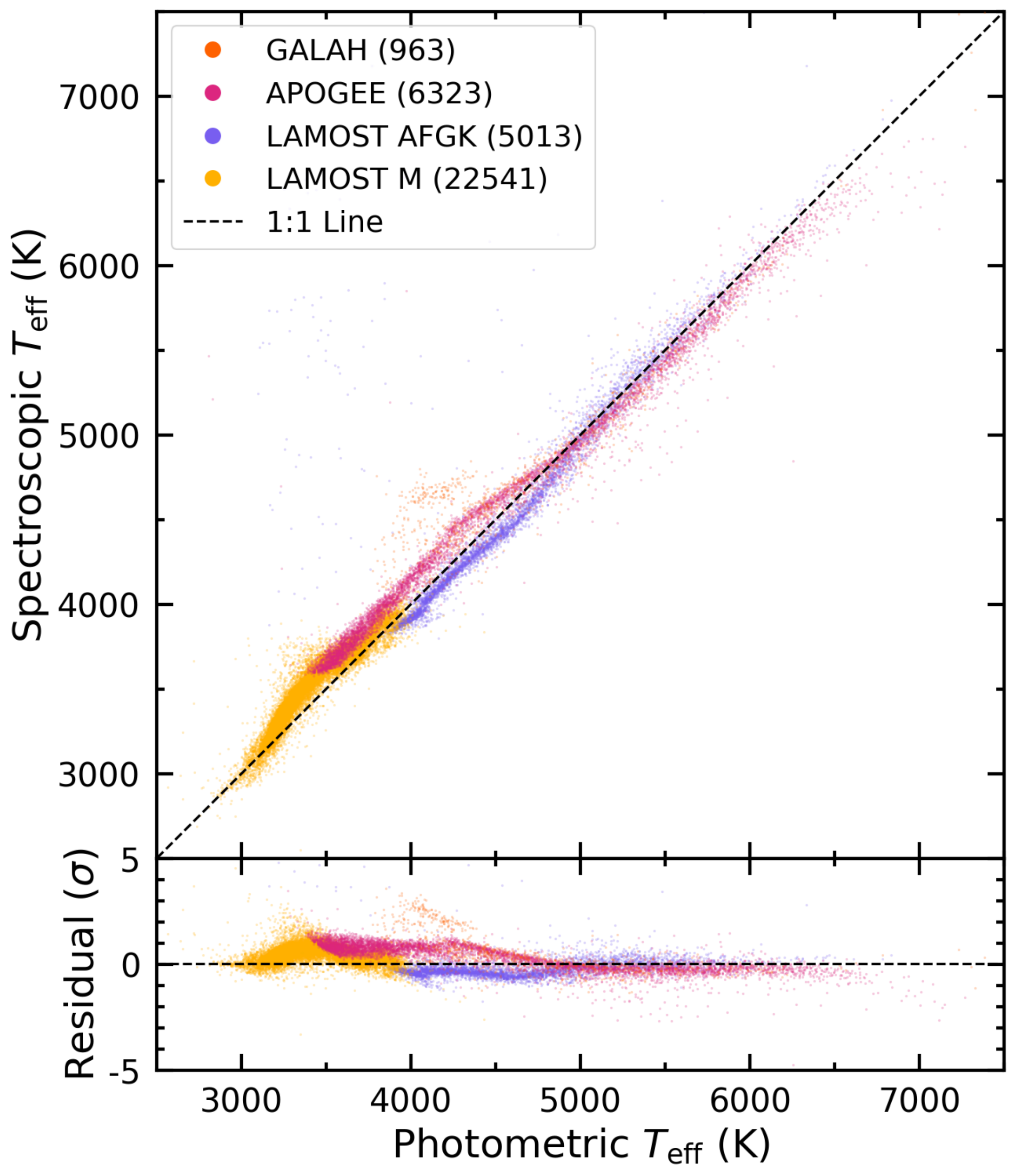}{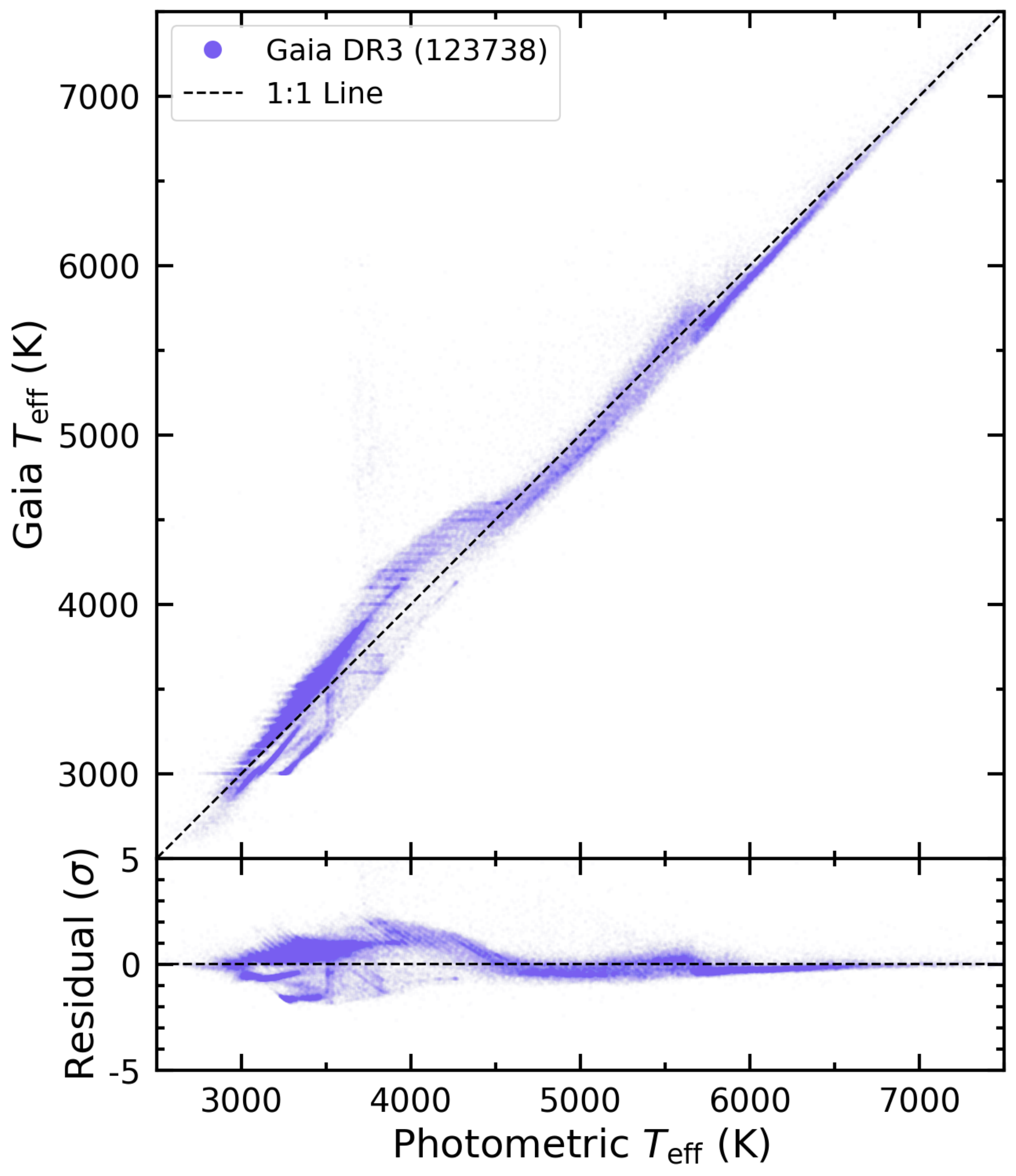}
\caption{(Left) Comparison of our photo-astrometric \teff\ values to spectroscopic \teff\ measurements from GALAH, APOGEE, and LAMOST. The number of targets that overlap between our catalog and the spectroscopic surveys is shown in the plot legend. Our measurements for cooler stars tend to be lower than all three spectroscopic surveys, with notable differences starting around 4,500~K for GALAH and APOGEE, and 3,500~K for LAMOST. (Right) Comparison of our temperatures to values in Gaia DR3. Our temperatures are typically systematically lower for stars below $\sim$4,500~K and higher above that value, which can be attributed to either molecular features in cool star spectra affecting \teff\ estimates, or a typical overestimation of extinction for nearby cool stars causing spectroscopic \teff\ measurements to be overestimated \citep{Andrae2018,Dressing2019}.} \label{fig:teffspec}
\end{figure*}

To check our derived stellar parameters, we first compared our \teff\ values to spectroscopically derived \teff\ measurements for common targets from the GALactic Archaeology with HERMES survey \citep[GALAH,][]{DeSilva2015,Buder2021}, the Apache Point Observatory Galactic Evolution Experiment survey \citep[APOGEE,][]{Majewski2017,Abdurro'uf2022}, and the Large Sky Area Multi-Object Fiber Spectroscopic Telescope survey \citep[LAMOST,][]{Cui2012,Zhao2012}. Figure~\ref{fig:teffspec} shows reasonable agreement between these \teff\ measurements, with 93\% of our \teff\ values within $1\sigma$ of the spectroscopic \teff\ measurements. Typical \teff\ differences are 2.9\% for APOGEE stars, 2.3\% for GALAH stars, 1.7\% for LAMOST AFGK stars and 3.4\% for LAMOST M stars, which indicates good agreement overall, and are on the order of our median \teff\ uncertainties of $\sim$3\%. We also compared our \teff\ measurements to those derived by the \gaia\ Apsis pipeline \citep{Creevey2022,Fouesneau2022} and found typical differences of 2.5\% (Figure~\ref{fig:teffspec}), with 86\% of our \teff\ values within $1\sigma$ of the \gaia\ measurements. Notably, for cool stars below $\sim$4500\,K, our photometric \teff\ values tend to be lower than the spectroscopic or \gaia\ measurements. This effect has been seen before \citep[e.g,][]{Andrae2018,Dressing2019} and can be attributed to either strong molecular features in cool star spectra affecting \teff\ estimates, or that extinction, and consequently \teff\, is typically overestimated for nearby cool stars because the spectroscopic training sample for stellar parameters is typically much further away.

\begin{figure*}[ht!]
\includegraphics[width=0.99\linewidth, trim=2 2 2 2, clip]{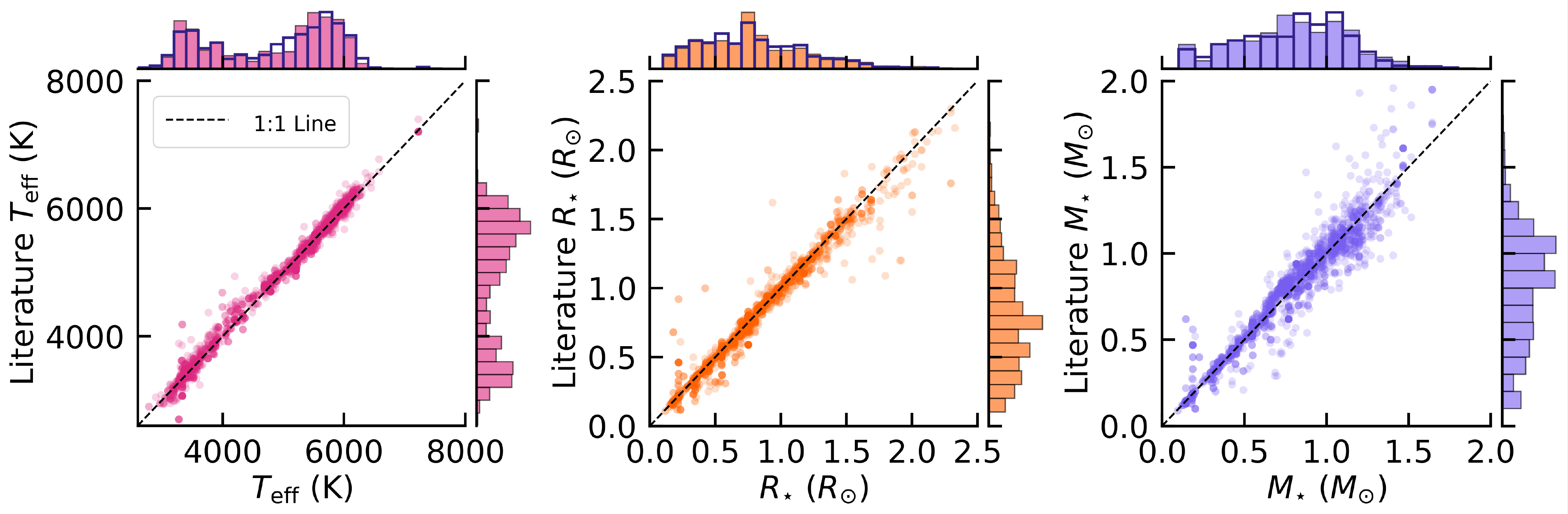}
\caption{Comparison of our \teff, $R_{\star}$, and $M_{\star}$ measurements to those of confirmed exoplanet systems from the literature. Above each plot is the histogram of our measured values, and to the right of each plot is a histogram of the literature measured values. For direct comparison, the dark blue histogram outlines of the literature values are plotted over our measured values. Compared to literature values, our measurements have average differences of 1\% in \teff\, 3\% in $R_{\star}$, and 5.5\% in $M_{\star}$. \label{fig:plcomp}}
\end{figure*}

Since we are interested in the suitability of stars to be hosts of potentially observable exoplanets, we also compared our catalog to measurements of confirmed exoplanet hosts from the NASA Exoplanet Archive (Figure~\ref{fig:plcomp}). We note that any given system can have multiple stellar measurements, so we took all existing \teff\, $R_{\star}$, and $M_{\star}$ measurements from the Planetary Systems table \citep{ps}. Our measurements are again in good agreement with the literature values, with average differences of 1\% in \teff, 3\% in $R_{\star}$, and 5.5\% in $M_{\star}$.

\section{Finding suitable exoplanets to search for atmospheric \texorpdfstring{O$_{\mathrm{2}}$}{O2}} \label{sec:o2}

With our homogeneous catalog of nearby stars, we can now switch gears to focus on exoplanets. We first require a suitable planet amenable to atmospheric \ot\ measurement with upcoming ELTs. For biosignature searches, we would like to find a transiting planet in a nearby system within the habitable zone of its host star.

Within 20\,pc, there are currently 31 confirmed planets discovered via the transit method, and all but three of them orbit M dwarfs \citep{PSCompPars}. However, only two of these systems contain planets in the habitable zone, \mbox{TRAPPIST-1} and LHS~1140, which we explore further in Section~\ref{sec:nhz}. Of note, the MEarth Project \citep{Nutzman2008} has surveyed nearly 3000 mid-to-late M dwarfs.\footnote{As of April 2023, \url{https://lweb.cfa.harvard.edu/MEarth/DataDR11.html}} However, MEarth is not typically sensitive to Earth-sized planets at habitable zone distances \citep{Berta2013}. Similarly, the $\sim$27 day observation duration for each TESS sector is not conducive to identifying habitable zone planets. For example, TESS observed the planet LP~890-9~b, a super-Earth orbiting an M6~V star every 2.73 days, and only after extensive follow-up observations was another super-Earth within the star's habitable zone (LP~890-9~c) identified \citep{Delrez2022}. Results from a transit injection-recovery test on LP~890-9 TESS data by \citet{Delrez2022} show that it is very unlikely that TESS is sensitive to habitable zone Earth-sized planets orbiting M dwarfs. There are some regions of the sky for which there are overlapping TESS sectors, and these might be the most promising regions to search for habitable zone Earths with existing survey data. In general, \citet{Brady2022} found that TESS has found fewer planets around M dwarfs than predicted from Kepler simulations using Kepler data \citep[e.g.,][]{Dressing2015,Barclay2018}, especially for mid-to-late M dwarfs. Therefore, Earth-sized transiting planets could still exist in the habitable zone of nearby stars, but have not yet been discovered. To estimate that number we refer to \ee.

\subsection{Accounting for \texorpdfstring{\ee}{eta-earth}}\label{sec:ee}

There is no current consensus on the value of \ee, the occurrence rate of Earth-sized planets in the habitable zone of stars. For Sun-like stars (FGK), we have not yet observed a sufficient population of Earth-like exoplanets in the habitable zone, so estimates of \ee\ require assumptions and extrapolations. Estimates of \ee\ at the lower end of the range are $0.053^{+0.070}_{-0.037}$ \citep[Model \#7,][]{Pascucci2019} and $0.055^{+0.011}_{-0.009}$ \citep{Neil2020}. Mid-range estimates of \ee\ are $0.11^{+0.07}_{-0.05}$ \citep{Kunimoto2020} and $0.094^{+0.034}_{-0.025}$ \citep{Bergsten2022}. Toward the upper range of literature \ee\ values, \citet{Mulders2018} measured $0.36\pm0.14$, \citet{Pascucci2019} measured $0.536\pm0.297$ (Model \#2), and \citet{Bryson2021} posited a range of \ee\ between $0.37^{+0.48}_{-0.21}$ and $0.60^{+0.90}_{-0.36}$ (Model \#1). It is important to note that these studies use different methodology and slightly different planet radius and period ranges. Additionally, \citet{Pascucci2019}, \citet{Neil2020}, and \citet{Bergsten2022}, motivated by the detection of the planet radius valley \citep{Fulton2017}, realized and accounted for the possibility that the close-in small-planet population likely differs from the small-planet population in the habitable zone.

Literature values of \ee\ for early-type M dwarfs are in-between the mid and upper ranges for FGK dwarfs, with \citet{Dressing2015} measuring \ee\ $=0.16^{+0.17}_{-0.07}$ and \citet{Pinamonti2022} measuring \ee\ $< 0.23$ from radial velocity data. Values for \ee\ for mid and late-type M dwarfs have not been measured, however, there have been measurements of planet occurrence rates for these stars. We first note that occurrence rate estimates for early-type M dwarfs are higher than for FGK stars \citep[e.g.,][]{Dressing2015,Mulders2015a}, and \citet{Mulders2015b} found that M dwarfs have 3.5 times more small (1~\re\ $< R_p <$ 2.8~\re) planets than FGK dwarfs. \citet{Hardegree-Ullman2019} found evidence for a continued increase in planet occurrence rates for mid-type M dwarfs ($\sim$M3\,V--M5.5\,V), but their estimates are based on a small sample of 13 planets from \textit{Kepler}. \citet{Sagear2020} and \citet{Sestovic2020} searched \textit{K2} data for planets around late-type M dwarfs ($\sim$M6\,V--M9.5\,V) and early brown dwarfs, but they were not able to identify any Earth-sized planets in the data, and could only place upper limits on occurrence rates for late-type M dwarfs to $\lesssim1$ planet per star. Similarly, \citet{Dietrich2023} conducted a volume-limited survey of northern late-M dwarfs within 15\,pc and did not identify any planets at orbital periods less than one day. Indeed, \citet{Mulders2021} posited from the pebble accretion planet formation model that the occurrence rates for super-Earths peak at early-type M dwarfs, but decrease toward later-type M dwarfs.

\begin{figure}[ht!]
\includegraphics[width=0.99\linewidth]{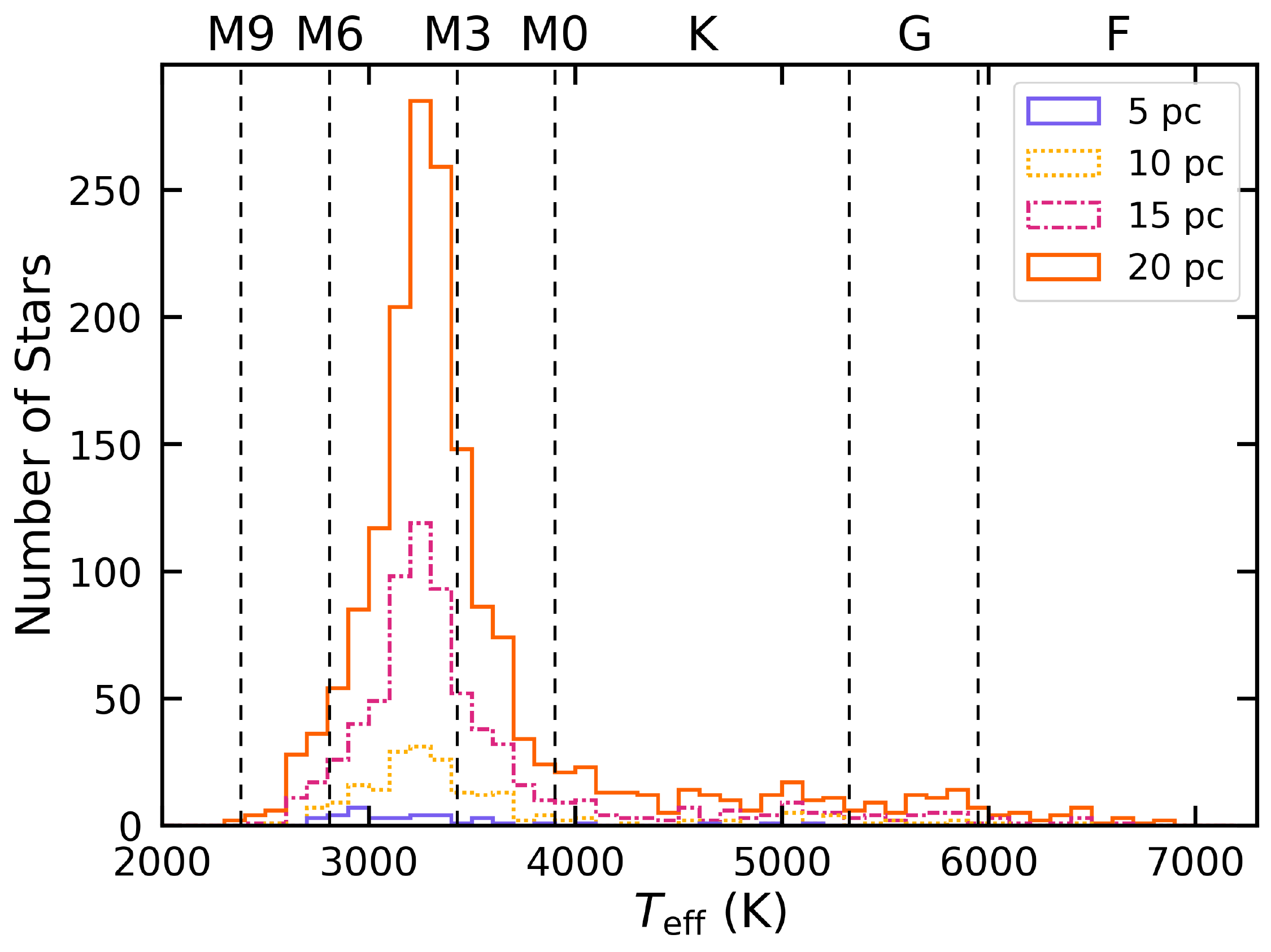}
\caption{The distribution of the nearby single-star FGKM dwarf population at intervals of 5\,pc out to 20\,pc. M dwarfs comprise a majority (84\%) of the nearby main sequence single-star population, with a peak around M4. \label{fig:20pcst}}
\end{figure}

Before we estimated the number of Earth-sized habitable zone planets observable with ELTs, we first imposed a limit to targets within 20\,pc. M dwarfs constitute a majority of the stellar population within 20\,pc (Figure~\ref{fig:20pcst}). Due to the paucity of FGK dwarfs within 20\,pc, along with their lower geometric transit probability and further out habitable zones, the contributions of these stars are negligible in the computation of observable nearby Earth-sized habitable zone planets. The limitation to 20\,pc is motivated by the fact that at further distances stars become fainter and more difficult to build up a sufficient transit signal. For example, a $G_{RP}=8.5$ mid-type M dwarf (the most common type of star within 20\,pc) at 5\,pc would be four times fainter ($G_{RP}=10$) at 10\,pc and 16 times fainter ($G_{RP}=11.5$) at 20\,pc, necessitating exposure times (and number of transit observations/years of observations) commensurate to the diminished brightness factor in order to achieve the same signal-to-noise.

\begin{figure*}[ht!]
\gridline{\includegraphics[width=0.48\textwidth, trim=2 2 2 2, clip]{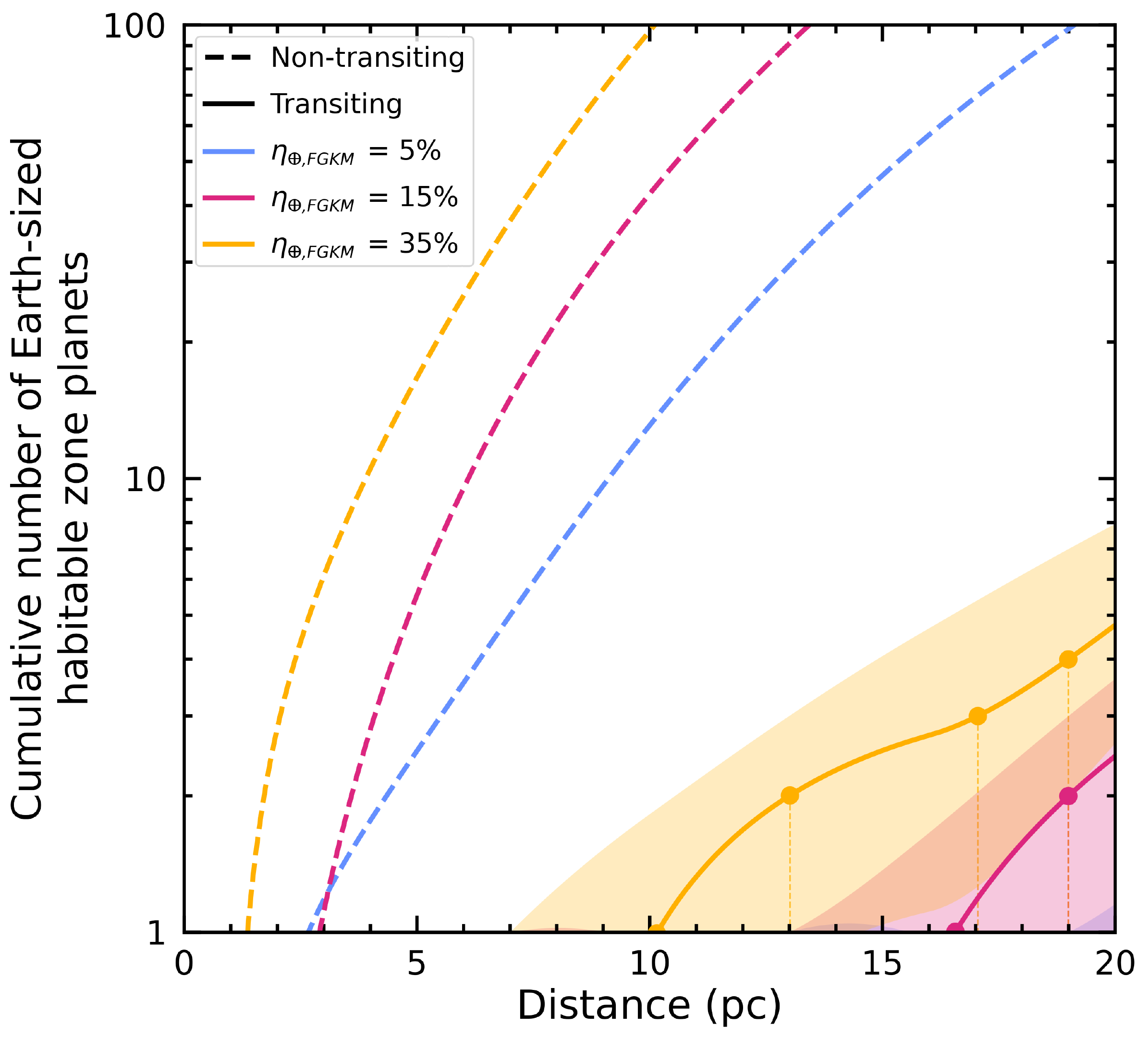}
          \includegraphics[width=0.48\textwidth, trim=2 2 2 2, clip]{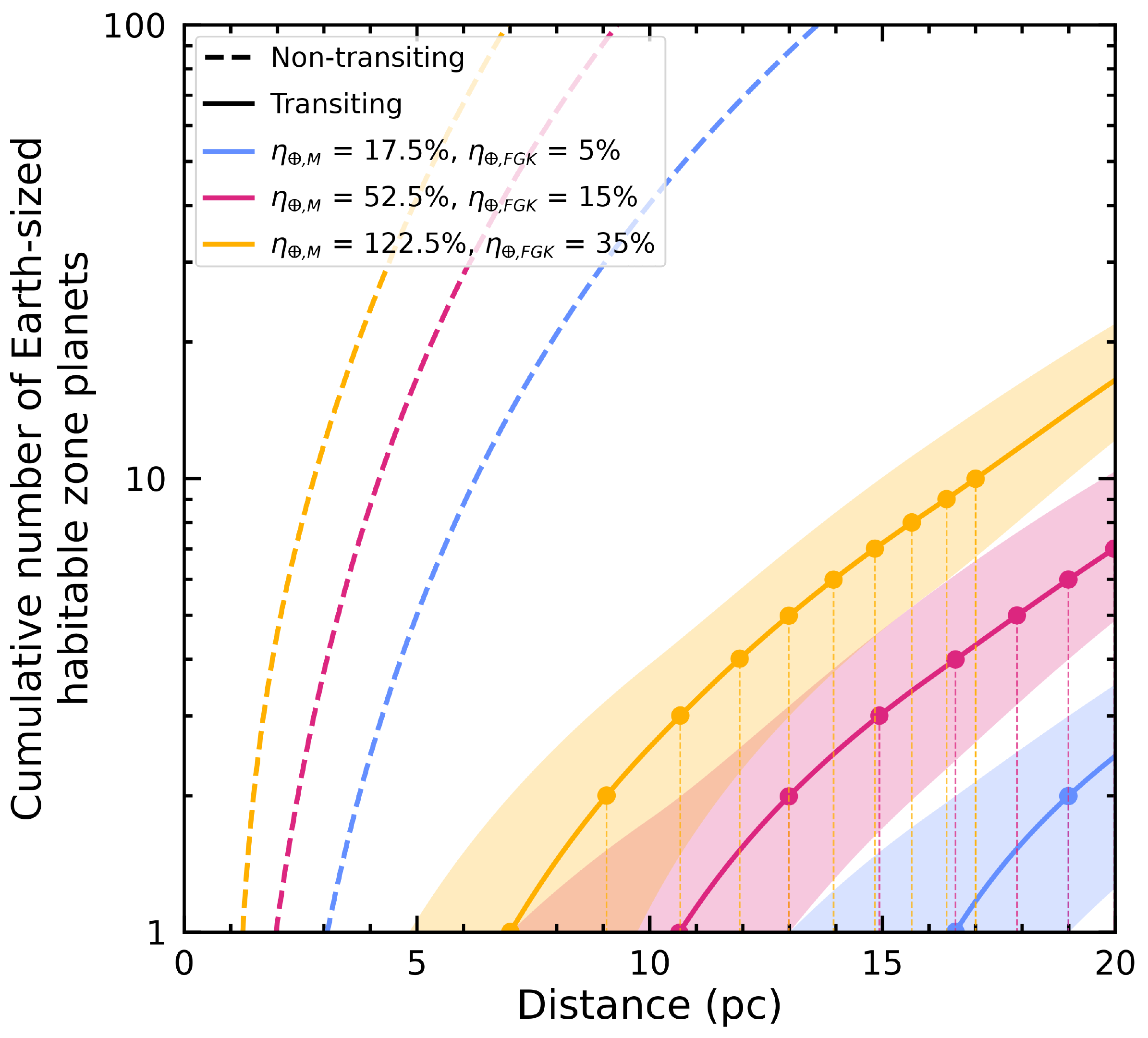}}
\vspace{-12pt}

\caption{Cumulative estimates of the number of Earth-sized habitable zone planets as a function of distance for planets in both transiting (solid lines) and non-transiting geometries (dashed lines). (Left) Scenario (1): \ee\ is the same for FGK and M dwarfs, assuming three different values for \ee: 5\%, 15\%, and 35\%. (Right) Scenario (2): \ee\ is 3.5 times higher for all M dwarfs than the baseline \ee\ values from (1). To aid the eye, we have added circles and extending vertical dashed lines on each of the transiting curves to show at which distance we expect to find one or more Earth-sized habitable zone planet. For these estimates, we used 1,739 single stars out to 20\,pc. There are notably significantly more non-transiting Earth-sized habitable zone planets, and within 10\,pc, we can expect to find at least 10 to 100 of these planets.}\label{fig:eta}
\end{figure*}

Given the above range of \ee\ values for FGK and early M stars, we considered two different scenarios assuming three different values for \ee. In the first scenario, we considered \ee\ = 0.05, 0.15, and 0.35, which are close to the median low, mid, and upper range of literature values of \ee\ for FGK and M stars. In the second scenario we considered a more optimistic value of \ee\ for M dwarfs, assuming a factor of 3.5 times higher \citep{Mulders2015b} than the baseline from the first scenario (\ee\ = 0.175, 0.525, and 1.225). We again used \texttt{Bioverse} to apply these \ee\ scenarios to the stars within 20\,pc, using the \citet{Kopparapu2014} conservative habitable zone boundaries between runaway and maximum greenhouse) and accounting for geometric transit probability to estimate both the transiting and non-transiting planet populations. In order to assess uncertainties, we generated 100 separate planet samples in \texttt{Bioverse} around all systems within 20\,pc of the Sun. For our simulations, we define Earth-sized as a planet with radius in the range $0.8S^{0.25} < R_p < 1.3\,R_{\oplus}$, where $S$ is stellar incident flux in units of Earth insolation flux. The lower limit is adopted from \citet{Bixel2021}, and is set by the theoretical minimum size a planet can be and still retain an atmosphere based on models from \citet{Zahnle2017}. The upper limit is set by the size for which a planet is likely to have thick hydrogen-rich atmosphere which can cause heavy blanketing of other molecules in the lower atmosphere \citep{Kimura2022}. We present the cumulative number of Earth-sized habitable zone planets as a function of distance in Figure~\ref{fig:eta}, showing the median, 16$^{\mathrm{th}}$, and 84$^{\mathrm{th}}$ percentiles for the expected transiting planet population. We note that additional scenarios, such as a pebble accretion model from \citet{Mulders2021} or a continuous increase in planet occurrence rates toward later M dwarfs from \citet{Hardegree-Ullman2019} yield similar results to our scenarios one and two, respectively. 

In both of our computed \ee\ scenarios, it is unlikely we will find a transiting Earth-like candidate within 5\,pc. Within 10\,pc, we expect to detect one or two transiting exoplanets, assuming an optimistic baseline \ee\ of at least 0.15. Out to 20\,pc, between $\sim$1 and 20 transiting, habitable zone, Earth-sized planets could exist, but a larger number of planets is contingent on optimistic \ee\ values, including a 3.5$\times$ boost for M dwarfs. It is worth noting that of the $\sim$200 single stars within 10\,pc we expect to find between 10 and 100 non-transiting Earth-sized habitable zone planets. Indeed, additional planets like Proxima Centauri~b \citep{Anglada-Escude2016} should be detectable with some current and next-generation radial velocity instruments around nearby, relatively quiescent stars with a large survey of the solar neighborhood.

\subsection{Habitable Zone Transits Observable from ELTs}\label{sec:hztr}

In Section~\ref{sec:ee} we determined that there could be one or more transiting habitable zone Earth-sized planets within 20\,pc, based on a range of \ee\ values, which could be suitable to probe for \ot. Before we account for \ee\ in a survey simulation, however, we first need to establish the observability of planets in the nearby stellar population from the ELTs. This can be done if we assume a hypothetical transiting Earth-sized habitable zone planet orbiting each star using \texttt{Bioverse}. We began with the 1,739 single stars within 20\,pc (Figure~\ref{fig:20pcst}). First, for each of these stars we adapted existing \texttt{Bioverse} functions to compute the \citet{Kopparapu2014} conservative habitable zone boundaries from stellar luminosity and \teff, the geometric transit probability $\mathcal{P}_{\mathrm{transit}} = (R_{\star}+R_{\oplus})/a$, considering an Earth-radius planet \re\ at semi-major axis $a$ in the middle of the star's habitable zone, the corresponding orbital period $P$ from Kepler's third law, and transit duration ($1^{\rm{st}}$ contact to $4^{\rm{th}}$ contact) $T_{\mathrm{14}} = (P/\pi)\sin^{-1}(\mathcal{P}_{\mathrm{transit}})$, assuming an ideal scenario where the impact parameter is zero. These parameters are listed in Table~\ref{tab:hzpl}.

\begin{deluxetable}{lccccc}[ht!]

\tablecaption{Habitable zone properties for M dwarfs within 20\,pc. The last three columns assume a $1\,R_{\oplus}$ planet in the middle of the star's habitable zone.\label{tab:hzpl}}

\tablehead{\colhead{Name} & \colhead{$a_{\mathrm{inner}}$} & \colhead{$a_{\mathrm{outer}}$} & \colhead{$\mathcal{P}_{\mathrm{transit}}$} & \colhead{$P$} & \colhead{$T_{\mathrm{14}}$}\vspace{-8pt} \\ 
\colhead{} & \colhead{(au)} & \colhead{(au)} & \colhead{(\%)} & \colhead{(d)} & \colhead{(hr)} } 

\startdata
Proxima Centauri & 0.038 & 0.077 & 1.27 & 14.64 & 1.42 \\
Barnard's star & 0.063 & 0.120 & 0.97 & 26.33 & 1.94 \\
Wolf 359 & 0.030 & 0.060 & 1.50 & 10.55 & 1.21 \\
HD 95735 & 0.157 & 0.306 & 0.80 & 66.91 & 4.10 \\
G 272-61B & 0.032 & 0.065 & 1.22 & 12.77 & 1.19 \\
G 272-61A & 0.041 & 0.082 & 1.29 & 15.34 & 1.52 \\
Ross 154 & 0.064 & 0.127 & 1.07 & 25.68 & 2.10 \\
Ross 248 & 0.045 & 0.090 & 1.17 & 17.94 & 1.60 \\
HD 217987 & 0.198 & 0.388 & 0.75 & 84.67 & 4.84 \\
Ross 128 & 0.061 & 0.120 & 1.12 & 23.54 & 2.02 \\
\enddata
\tablecomments{Table~\ref{tab:hzpl} is published in its entirety in machine-readable format. A small portion is shown here for guidance regarding its form and content.}
\vspace{-22pt}
\end{deluxetable}
\vspace{-12pt}

Next, we created a custom function within \texttt{Bioverse}, incorporating routines from \texttt{astropy} \citep{astropy:2013,astropy:2018,AstropyCollaboration2022} and \texttt{astroplan} \citep{astroplan2018} to simulate how many transits per year would be observable at night (between nautical twilight when the Sun is $12^{\circ}$ below the horizon) and with an airmass $<2$ from each of the three upcoming ELTs, again assuming each nearby star hosted a transiting Earth-sized planet within its conservative habitable zone. We assumed a one hour baseline on each side of the transit to be considered a full transit. We ran a Monte Carlo simulation 100 times on each star, selecting a different Earth-sized planet radius and habitable zone location from a log-uniform random distribution \citep[based on the \textit{Kepler} exoplanet semi-major axis distribution, e.g.,][]{Zeng2018} each iteration. We then computed the average total number of full transits per year and the average number of observable transits at the GMT, TMT, and E-ELT, again, under the assumption that each star hosts a transiting Earth-sized habitable zone planet. 

The average number of observable transits at each telescope location, however, is an upper limit. In order to observe a biosignature like \ot\ on an exoplanet from a ground-based telescope through our own \ot\ containing atmosphere, the relative velocity of a system with respect to Earth ($\nu_{\star}$) must be taken into account. For observations of the \ot\ {\it A} band at 760~nm, significant line blending above the average background blending fraction occurs at relative velocities of $\pm$13\,km~s$^{-1}$ and between about $\pm$(30 to 55)\,km~s$^{-1}$. For the near-infrared \ot\ band at 1268~nm blending zones of avoidance are at $\pm$4\,km~s$^{-1}$ and between about $\pm$(68 to 83)\,km~s$^{-1}$ \citep[see Figure~6 of][]{Lopez-Morales2019}. In order to account for line blending, we used the following equation from \citet{Lopez-Morales2019} to compute relative velocity:
\begin{equation}
    \nu_{\star} = \nu_{\mathrm{Sun},\star} + \nu_{\mathrm{Earth}} \cdot \sin\theta \cdot \cos\phi,
\end{equation}
where $\nu_{\mathrm{Sun},\star}$ is the stellar systemic velocity (Section~\ref{sec:sv}), $\nu_{\mathrm{Earth}}=29.78$\,km~s$^{-1}$ is Earth's velocity around the Sun, $\theta$ is the angle of the location of Earth along the ecliptic at the time of observation, and $\phi$ is the ecliptic latitude of the exoplanet host. We note that the orbital velocity of the exoplanet is not relevant because its radial component will be zero with respect to the host star during the observations since these occur during a transit. We considered three relative velocity scenarios of observability for (1) the \ot\ {\it A} band, (2) the \ot\ near-infrared band, and (3) when \ot\ is observable in both bands, removing transit events where the relative system velocity falls within the ranges where line blending occurs listed above. Scenario (3) requires the relative system velocity to be outside the range of both \ot\ A and near-infrared band blending.

An additional unavoidable burden to ground-based observers is weather. To get more realistic values of total observable transits per year from the ELTs, we applied a weather correction by multiplying the above values of transits per year by the average fraction of clear weather nights at each ELT site $N/365$, where $N$ is 300 for the GMT\footnote{Las Campanas, Chile: \url{https://www.cfa.harvard.edu/facilities-technology/telescopes-instruments/magellan-telescopes}}, 315 for the TMT\footnote{Assuming the telescope will be built on Maunakea, Hawaii: \url{https://maunakea.com/faq/}.}, and 320 for the E-ELT\footnote{Cerro Armazones, Chile: \url{https://www.eso.org/sci/facilities/eelt/site/}}.

\begin{figure*}[ht!]
\includegraphics[width=0.96\linewidth, trim=2 2 2 2, clip]{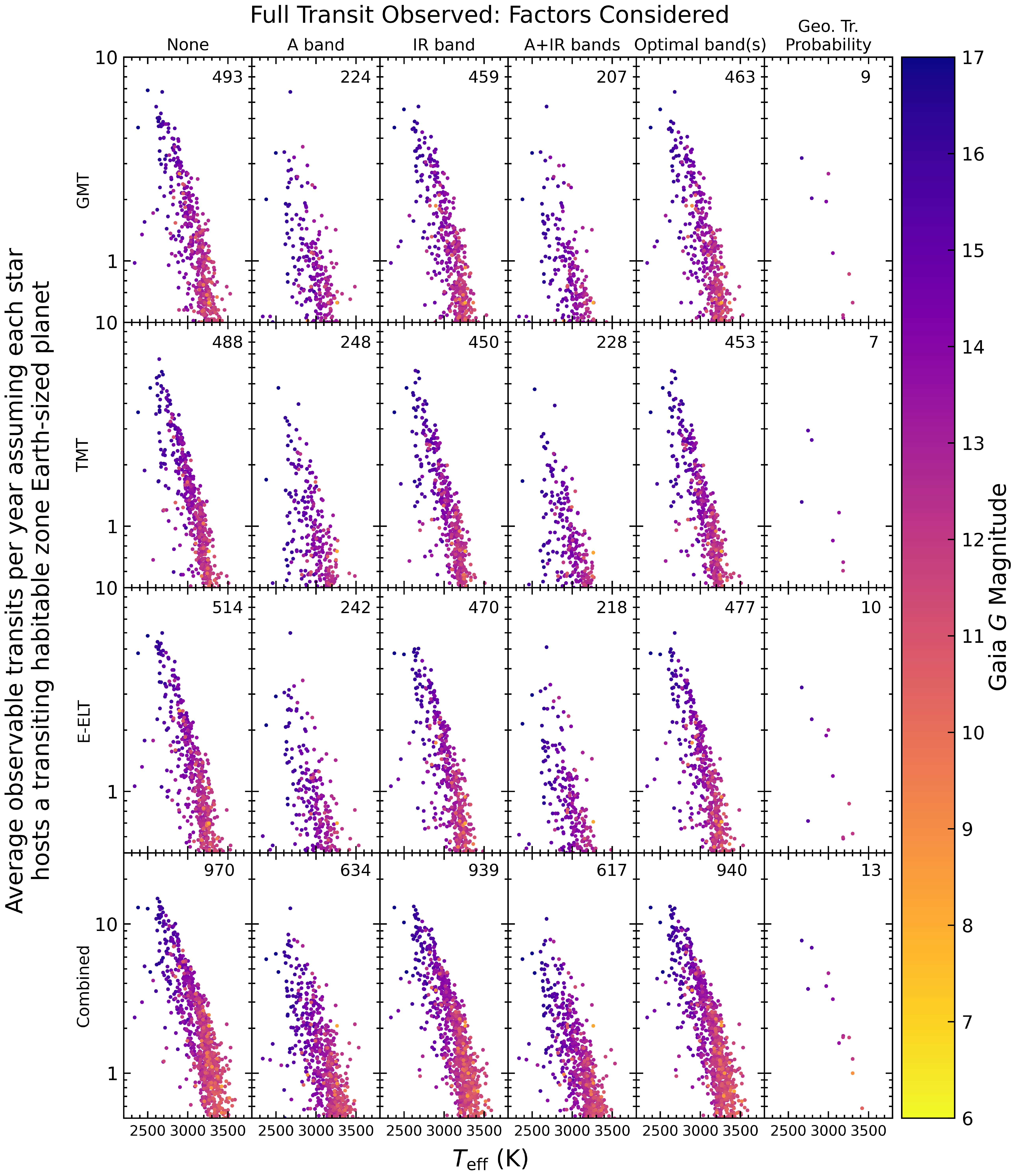}
\caption{Assuming each star hosts a transiting habitable-zone Earth-sized planet amenable to \ot\ observations from ground-based ELTs, this plot shows the average number of observable transits per year as a function of stellar effective temperature. The top three rows show observability from the three different ELTs, and the bottom row shows the combined total number of transits computed from the sum of the values for each star from each ELT. Each column shows different observability factors considered including no accounting for relative system velocity (None), accounting for system relative velocity at the time of observation for separate A and IR bands, when both A and IR are observable, optimal \ot\ band (maximum of A, IR, or A+IR), and geometric transit probability. The last column is a random sub-sample of 13 targets, which was selected based on the median geometric transit probability of 1.27\% of the 970 targets with the number of observable transits (from the first column) greater than 0.5. The total number of targets for each scenario is indicated in the upper right corner of each cell. Points are colored by apparent \gaia\ \textit{G}-band magnitude. \label{fig:o2obs}}
\end{figure*}

Our results for the average number of full observable transits per year, assuming a transiting habitable zone Earth-sized planet around each star and accounting for relative velocities and site weather, are available in Table~\ref{tab:fulltr} and visualized in Figure~\ref{fig:o2obs}. We highlight that our simulation shows no stars with spectral type earlier than $\sim$M2.5~V to be likely to have an average of at least one visible transit per year from the ground-based ELTs. This would make it very hard to build up enough signal to measure Earth-like levels of \ot\ on a planet orbiting an earlier-type star within a reasonable survey length.

\begin{figure*}[ht!]
\includegraphics[width=0.96\linewidth, trim=2 2 2 2, clip]{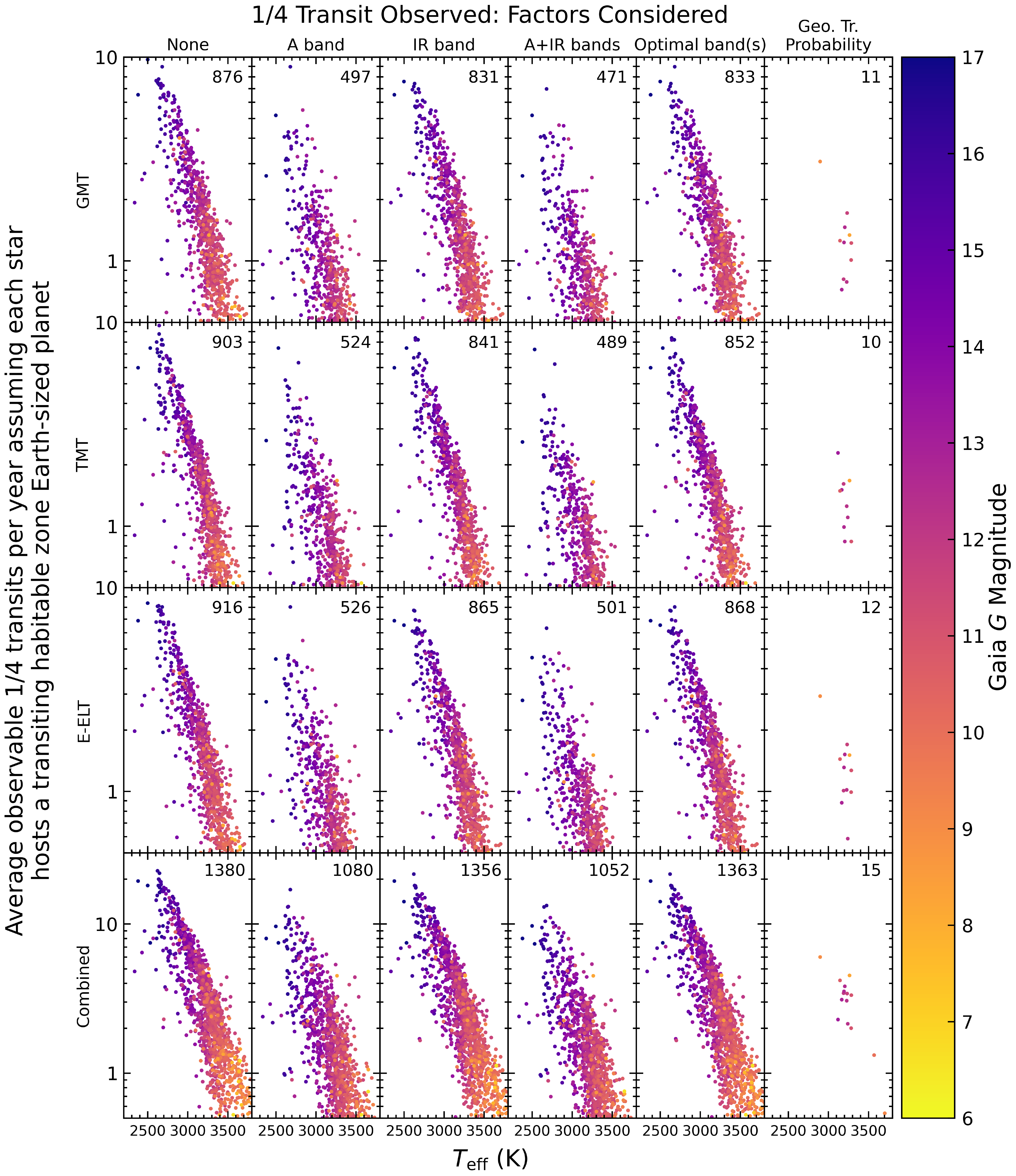}
\caption{Same as Figure~\ref{fig:o2obs}, but only requiring 1/4 of the full transit and baseline to be observable in order to establish an out-of-transit baseline and a transit depth. The last column is a random sub-sample of 16 targets (from 1,380 targets with the number of observable 1/4 transits per year greater than 0.5 and a median transit probability 1.15\%). \label{fig:o2obsq}}
\end{figure*}

We also considered a more optimistic scenario in which we do not require a full transit and baseline, since we technically only need to measure the difference between in and out-of-transit depth of the \ot\ lines. For simplicity, we ran the same simulation as above, but only requiring one quarter of a full transit and baseline to be observable, which would encompass the time of a partial baseline and in-transit observation. The results from this simulation are presented in Table~\ref{tab:parttr} and Figure~\ref{fig:o2obsq}, showing on average two times more observable transit events, including a few early-type M dwarfs with one or two observable events per year.

\begin{deluxetable}{lc|ccccchhhhhhhhhh}[ht!]
\tablecaption{Simulated average number of visible \emph{full} transits per year from each ELT for a hypothetical habitable zone Earth-sized planet orbiting M dwarfs within 20\,pc. For each ELT, we consider the number of full transits visible from the ground per year with no relative system velocity requirements (None), considering relative system velocities required to observe in the \ot\ A and IR bands separately, when relative system velocities allow both A and IR band observations, and the best case \ot\ band scenario, which is the maximum of the A, IR, and A+IR columns. Three dots indicate our simulations yielded no observable transits for that particular scenario.}\label{tab:fulltr}

\tablehead{\colhead{Name} & \colhead{Transits} & \multicolumn{5}{c}{GMT}\vspace{-8pt} \\%& \multicolumn{5}{c}{TMT} & \multicolumn{5}{c}{E-ELT} \\
\colhead{} & \colhead{per yr} & \colhead{None} & \colhead{A} & \colhead{IR} & \colhead{A+IR} & \colhead{Best} & \nocolhead{None} & \nocolhead{A} & \nocolhead{IR} & \nocolhead{A+IR} & \nocolhead{Best} & \nocolhead{None} & \nocolhead{A} & \nocolhead{IR} & \nocolhead{A+IR} & \nocolhead{Best} }

\startdata
Proxima Centauri & 26 & 2.68 & 0.60 & 1.87 & 0.60 & 1.87 & \nodata & \nodata & \nodata & \nodata & \nodata & 2.49 & 0.47 & 1.73 & 0.47 & 1.73 \\
Barnard's star & 15 & 0.62 & 0.62 & 0.62 & 0.62 & 0.62 & 0.75 & 0.75 & 0.75 & 0.74 & 0.75 & 0.70 & 0.70 & 0.70 & 0.71 & 0.71 \\
Wolf 359 & 38 & 2.47 & 0.37 & 1.87 & 0.37 & 1.87 & 2.74 & 0.58 & 2.50 & 0.57 & 2.50 & 2.29 & 0.42 & 2.16 & 0.43 & 2.16 \\
HD 95735 & 5 & \nodata & \nodata & \nodata & \nodata & \nodata & 0.11 & 0.11 & 0.88 & 0.86 & 0.88 & \nodata & \nodata & \nodata & \nodata & \nodata \\
G 272-61B & 33 & 2.38 & 0.75 & 2.14 & 0.75 & 2.14 & 1.78 & 0.60 & 1.58 & 0.59 & 1.58 & 2.49 & 0.78 & 2.18 & 0.79 & 2.18 \\
G 272-61A & 26 & 1.54 & 0.30 & 1.32 & 0.30 & 1.32 & 1.36 & 0.24 & 1.78 & 0.23 & 1.78 & 1.60 & 0.33 & 1.35 & 0.37 & 1.35 \\
Ross 154 & 15 & 1.27 & 0.37 & 0.93 & 0.37 & 0.93 & 0.39 & 0.11 & 0.33 & 0.11 & 0.33 & 1.44 & 0.37 & 0.95 & 0.38 & 0.95 \\
Ross 248 & 22 & \nodata & \nodata & \nodata & \nodata & \nodata & 1.65 & 1.65 & 1.26 & 1.24 & 1.65 & \nodata & \nodata & \nodata & \nodata & \nodata \\
HD 217987 & 4 & 0.13 & 0.42 & 0.12 & 0.42 & 0.42 & \nodata & \nodata & \nodata & \nodata & \nodata & 0.13 & 0.43 & 0.11 & 0.44 & 0.44 \\
Ross 128 & 17 & 0.76 & 0.99 & 0.65 & 0.99 & 0.99 & 1.35 & 0.18 & 0.95 & 0.17 & 0.95 & 0.83 & 0.14 & 0.72 & 0.15 & 0.72 \\
\enddata

\tablecomments{Table~\ref{tab:fulltr} is published in its entirety in machine-readable format with additional columns for the TMT and E-ELT. A small portion is shown here for guidance regarding its form and content.}

\end{deluxetable}
\vspace{-12pt}

\begin{deluxetable}{lc|ccccchhhhhhhhhh}[ht!]
\tablecaption{Simulated average number of visible \emph{partial} transits per year from each ELT for a hypothetical habitable zone Earth-sized planet orbiting M dwarfs within 20\,pc. For each ELT, we consider the number of partial transits visible from the ground per year with no relative system velocity requirements (None), considering relative system velocities required to observe in the \ot\ A and IR bands separately, when relative system velocities allow both A and IR band observations, and the best case \ot\ band scenario, which is the maximum of the A, IR, and A+IR columns. Three dots indicate our simulations yielded no observable transits for that particular scenario.}\label{tab:parttr}

\tablehead{\colhead{Name} & \colhead{Transits} & \multicolumn{5}{c}{GMT}\vspace{-8pt} \\%& \multicolumn{5}{c}{TMT} & \multicolumn{5}{c}{E-ELT} \\
\colhead{} & \colhead{per yr} & \colhead{None} & \colhead{A} & \colhead{IR} & \colhead{A+IR} & \colhead{Best} & \nocolhead{None} & \nocolhead{A} & \nocolhead{IR} & \nocolhead{A+IR} & \nocolhead{Best} & \nocolhead{None} & \nocolhead{A} & \nocolhead{IR} & \nocolhead{A+IR} & \nocolhead{Best} }

\startdata
Proxima Centauri & 27 & 4.20 & 1.14 & 3.74 & 1.14 & 3.74 & \nodata & \nodata & \nodata & \nodata & \nodata & 3.80 & 1.96 & 2.93 & 1.11 & 2.93 \\
Barnard's star & 15 & 1.34 & 1.34 & 1.34 & 1.34 & 1.34 & 1.67 & 1.67 & 1.67 & 1.65 & 1.67 & 1.48 & 1.48 & 1.48 & 1.58 & 1.58 \\
Wolf 359 & 41 & 3.53 & 0.85 & 3.19 & 0.85 & 3.19 & 4.93 & 1.15 & 4.49 & 1.87 & 4.49 & 3.89 & 0.89 & 3.51 & 0.93 & 3.51 \\
HD 95735 & 5 & \nodata & \nodata & \nodata & \nodata & \nodata & 0.53 & 0.53 & 0.46 & 0.45 & 0.53 & \nodata & \nodata & \nodata & \nodata & \nodata \\
G 272-61B & 32 & 3.68 & 1.63 & 3.21 & 1.63 & 3.21 & 3.22 & 1.55 & 2.85 & 1.53 & 2.85 & 3.87 & 1.71 & 3.38 & 1.73 & 3.38 \\
G 272-61A & 26 & 3.14 & 0.79 & 2.55 & 0.79 & 2.55 & 2.32 & 0.52 & 1.89 & 0.59 & 1.89 & 3.37 & 0.84 & 2.72 & 0.85 & 2.72 \\
Ross 154 & 16 & 1.86 & 0.85 & 1.68 & 0.85 & 1.68 & 1.22 & 0.58 & 1.12 & 0.57 & 1.12 & 1.93 & 0.85 & 1.74 & 0.86 & 1.74 \\
Ross 248 & 21 & \nodata & \nodata & \nodata & \nodata & \nodata & 2.75 & 2.64 & 2.22 & 2.71 & 2.64 & \nodata & \nodata & \nodata & \nodata & \nodata \\
HD 217987 & 4 & 0.53 & 0.34 & 0.49 & 0.34 & 0.49 & 0.14 & 0.79 & 0.14 & 0.78 & 0.14 & 0.54 & 0.34 & 0.49 & 0.34 & 0.49 \\
Ross 128 & 16 & 1.54 & 0.21 & 1.39 & 0.21 & 1.39 & 1.64 & 0.37 & 1.45 & 0.33 & 1.45 & 1.62 & 0.23 & 1.49 & 0.24 & 1.49 \\
\enddata

\tablecomments{Table~\ref{tab:parttr} is published in its entirety in machine-readable format with additional columns for the TMT and E-ELT. A small portion is shown here for guidance regarding its form and content.}

\end{deluxetable}
\vspace{-12pt}

\section{Timeline for surveying for Earth-like \texorpdfstring{O$_{\mathrm{2}}$}{O2} concentrations} \label{sec:time}

Assuming a suitable planet to search for atmospheric \ot\ exists, we still need to determine how long it will take to survey the planet for Earth-like \ot\ concentrations. This depends not only on the properties of the target system (e.g., host star spectral type, distance, planet size), but also on telescope and instrument systematics and detection thresholds (e.g., 3$\sigma$, 5$\sigma$, 7$\sigma$).

\begin{figure*}[ht!]
\includegraphics[width=0.99\linewidth, trim=2 2 2 2, clip]{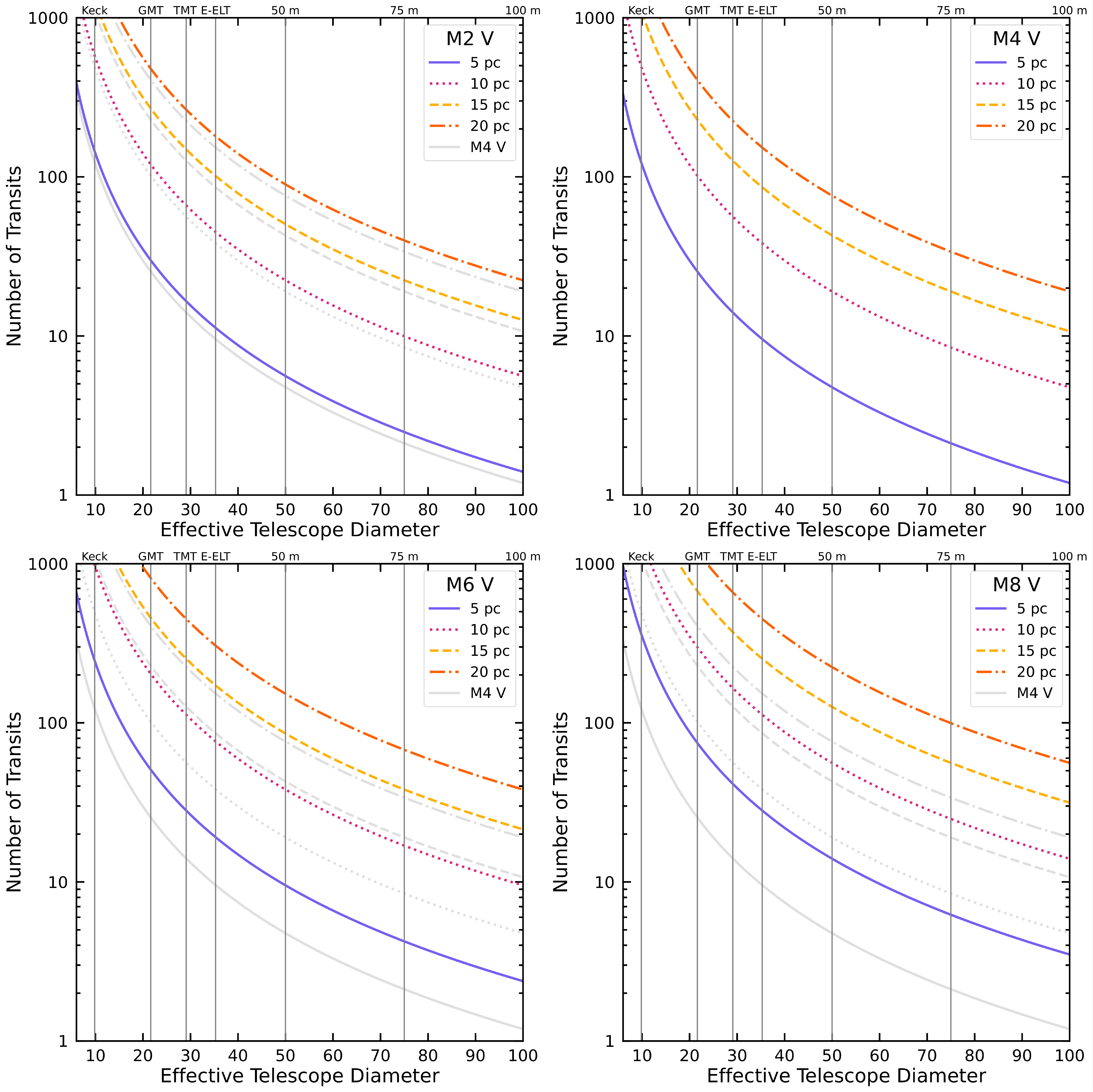}
\caption{Estimated number of transits required to probe for Earth-like \ot\ concentrations at 3$\sigma$ significance on an Earth-sized habitable zone planet orbiting M dwarfs as a function of telescope diameter. Shown are the models for hypothetical planets orbiting M2~V, M4~V, M6~V, and M8~V stars at distances of 5 (blue solid), 10 (magenta dotted), 15 (yellow dashed), and 20\,pc (red dashed-dotted). These models show that a hypothetical Earth-sized habitable zone planet orbiting an M4~V star would require the fewest number of transits to probe for Earth-like concentrations of \ot. In the M2~V, M6~V, and M8~V plots we show the M4~V models in light gray for reference. Vertical dark gray lines denote the effective diameters (based on total collecting area) of the Keck \citep[9.84~m,][]{Mast1986}, GMT \citep[21.65~m,][]{Jacoby2012}, TMT \citep[29.08~m,][]{Skidmore2015}, E-ELT \citep[35.29~m,][]{Liske2012}, and theoretical 50, 75, and 100~m diameter equivalent telescopes. \label{fig:transit_models}}
\end{figure*}

For our calculations, we used the models generated by \citet{Lopez-Morales2019}, which determined the number of transits required for a 3$\sigma$ detection of Earth-like \ot\ concentrations on hypothetical transiting Earth analogs, accounting for white noise, and assuming an instrument similar to G-CLEF on the GMT with a resolving power $R\approx100,000$ \citep{Szentgyorgyi2012}. For more details about these models, we refer the reader to Section 3.2 of \citet{Lopez-Morales2019}. The simulations presented in this section are based on the nominal resolution for currently planned first-generation instruments on the ELTs, however, in Appendix~\ref{app:500k} we include the same simulations for a hypothetical $R=500,000$ spectrograph. In Figure~\ref{fig:transit_models}, we show the number of planet transits required to detect Earth-like \ot\ concentrations at 3$\sigma$ significance versus telescope diameter for M2~V, M4~V, M6~V, and M8~V stars at 5, 10, 15, and 20\,pc. We highlighted the number of transits for each of the ELTs, and extended the models out to a hypothetical telescope diameter of 100-meters.

We interpolated these models using the distance of each of our targets, and the average number of observable transits per year from Section~\ref{sec:hztr} to compute the number of years it would take to test for Earth-like \ot\ levels on hypothetical nearby habitable zone Earth-sized planets using each ELT. We also computed the time to test for Earth-like \ot\ levels on these simulated planets by combining data from the ELTs, and for the hypothetical scenario in which we had ground-based telescopes with apertures of 50, 75, and 100-meters in diameter with a similar spectrograph to the ELTs. These calculations for hypothetical Earth analogs orbiting each nearby M dwarf are listed in Table~\ref{tab:fullyr}, assuming full transits are required, and in Table~\ref{tab:partyr}, assuming partial transits are required.

\begin{deluxetable}{lhhhhhhhhh|c|c|c|c}

\tablecaption{Time span (in years) required to probe for Earth-like \ot\ levels at 3$\sigma$ significance on a hypothetical transiting habitable zone Earth analog orbiting M dwarfs within 20\,pc (\emph{full} transits).\label{tab:fullyr}}
\tablehead{\colhead{Name} & \nocolhead{} & \nocolhead{} & \nocolhead{} & \nocolhead{} & \nocolhead{} & \nocolhead{} & \nocolhead{} & \nocolhead{} & \nocolhead{} & \colhead{Combined} & \colhead{50-m} & \colhead{75-m} & \colhead{100-m}\vspace{-8pt} \\ %\multicolumn{3}{c}{GMT} & \multicolumn{3}{c}{TMT} & \multicolumn{3}{c}{E-ELT} &
\colhead{} & \nocolhead{A} & \nocolhead{IR} & \nocolhead{A+IR} & \nocolhead{A} & \nocolhead{IR} & \nocolhead{A+IR} & \nocolhead{A} & \nocolhead{IR} & \nocolhead{A+IR} & \colhead{ELTs} & \colhead{} & \colhead{} & \colhead{} } 
\startdata
Proxima Centauri & 12.05 & 3.29 & 12.05 & \nodata & \nodata & \nodata & 4.90 & 1.32 & 4.90 & 0.94 & 0.62 & 0.28 & 0.16 \\
Barnard's star & 12.50 & 12.50 & 12.50 & 5.93 & 5.93 & 5.93 & 4.16 & 4.16 & 4.16 & 2.04 & 1.98 & 0.88 & 0.50 \\
Wolf 359 & 62.42 & 12.37 & 62.42 & 22.85 & 5.29 & 22.85 & 20.31 & 4.08 & 20.31 & 1.94 & 1.77 & 0.79 & 0.44 \\
HD 95735 & \nodata & \nodata & \nodata & 76.22 & 99.08 & 99.08 & \nodata & \nodata & \nodata & 76.22 & 25.46 & 11.32 & 6.37 \\
G 272-61B & 33.56 & 11.93 & 33.56 & 24.11 & 9.53 & 24.11 & 12.02 & 4.28 & 12.02 & 2.37 & 2.13 & 0.95 & 0.53 \\
G 272-61A & 98.86 & 22.24 & 98.86 & 70.78 & 15.54 & 70.78 & 36.03 & 8.03 & 36.03 & 4.28 & 4.00 & 1.78 & 1.00 \\
Ross 154 & 54.23 & 21.60 & 54.23 & 100.79 & 34.48 & 100.79 & 20.11 & 7.93 & 20.11 & 4.97 & 3.95 & 1.76 & 0.99 \\
Ross 248 & \nodata & \nodata & \nodata & 12.13 & 15.83 & 15.83 & \nodata & \nodata & \nodata & 12.13 & 4.05 & 1.80 & 1.01 \\
HD 217987 & 513.16 & 183.27 & 513.16 & \nodata & \nodata & \nodata & 181.80 & 69.92 & 181.80 & 50.61 & 34.52 & 15.34 & 8.63 \\
Ross 128 & 225.50 & 34.25 & 225.50 & 72.65 & 13.45 & 72.65 & 79.89 & 11.55 & 79.89 & 5.26 & 4.49 & 2.00 & 1.12 \\
\enddata

\tablecomments{Table~\ref{tab:fullyr} is published in its entirety in machine-readable format with additional columns for the GMT, TMT, and E-ELT in different \ot\ bands. A small portion is shown here for guidance regarding its form and content.}
\vspace{-32pt}
\end{deluxetable}
\vspace{-12pt}

\begin{deluxetable}{lhhhhhhhhh|c|c|c|c}

\tablecaption{Time span (in years) required to test for Earth-like \ot\ levels at 3$\sigma$ significance on a hypothetical transiting habitable zone Earth analog orbiting M dwarfs within 20\,pc (\emph{partial} transits).\label{tab:partyr}}
\tablehead{\colhead{Name} & \nocolhead{} & \nocolhead{} & \nocolhead{} & \nocolhead{} & \nocolhead{} & \nocolhead{} & \nocolhead{} & \nocolhead{} & \nocolhead{} & \colhead{Combined} & \colhead{50-m} & \colhead{75-m} & \colhead{100-m}\vspace{-8pt} \\ %\multicolumn{3}{c}{GMT} & \multicolumn{3}{c}{TMT} & \multicolumn{3}{c}{E-ELT} &
\colhead{} & \nocolhead{A} & \nocolhead{IR} & \nocolhead{A+IR} & \nocolhead{A} & \nocolhead{IR} & \nocolhead{A+IR} & \nocolhead{A} & \nocolhead{IR} & \nocolhead{A+IR} & \colhead{ELTs} & \colhead{} & \colhead{} & \colhead{} } 
\startdata
Proxima Centauri & 5.38 & 2.00 & 5.38 & \nodata & \nodata & \nodata & 2.08 & 0.78 & 2.08 & 0.56 & 0.38 & 0.17 & 0.09 \\
Barnard's star & 5.83 & 5.83 & 5.83 & 2.67 & 2.67 & 2.67 & 1.96 & 1.96 & 1.96 & 0.95 & 0.89 & 0.40 & 0.22 \\
Wolf 359 & 28.66 & 7.26 & 28.66 & 11.97 & 2.95 & 11.97 & 9.66 & 2.45 & 9.66 & 1.13 & 0.98 & 0.44 & 0.25 \\
HD 95735 & \nodata & \nodata & \nodata & 16.51 & 19.05 & 19.05 & \nodata & \nodata & \nodata & 16.51 & 5.52 & 2.45 & 1.38 \\
G 272-61B & 15.66 & 7.81 & 15.66 & 9.26 & 5.06 & 9.26 & 5.49 & 2.76 & 5.49 & 1.45 & 1.38 & 0.61 & 0.34 \\
G 272-61A & 37.07 & 11.48 & 37.07 & 32.39 & 8.85 & 32.39 & 13.00 & 4.00 & 13.00 & 2.22 & 2.00 & 0.89 & 0.50 \\
Ross 154 & 24.90 & 11.96 & 24.90 & 19.85 & 10.24 & 19.85 & 8.82 & 4.28 & 8.82 & 2.41 & 2.13 & 0.95 & 0.53 \\
Ross 248 & \nodata & \nodata & \nodata & 7.57 & 9.01 & 9.50 & \nodata & \nodata & \nodata & 7.57 & 2.53 & 1.12 & 0.63 \\
HD 217987 & 62.58 & 42.76 & 62.58 & 153.08 & 86.11 & 153.08 & 23.31 & 15.95 & 23.31 & 10.23 & 7.95 & 3.53 & 1.99 \\
Ross 128 & 104.08 & 16.01 & 104.08 & 41.51 & 8.81 & 41.51 & 35.51 & 5.54 & 35.51 & 2.81 & 2.76 & 1.23 & 0.69 \\
\enddata

\tablecomments{Table~\ref{tab:partyr} is published in its entirety in machine-readable format with additional columns for the GMT, TMT, and E-ELT in different \ot\ bands. A small portion is shown here for guidance regarding its form and content.}
\vspace{-32pt}
\end{deluxetable}
\vspace{-12pt}

\begin{figure*}[ht!]
\includegraphics[width=0.98\linewidth]{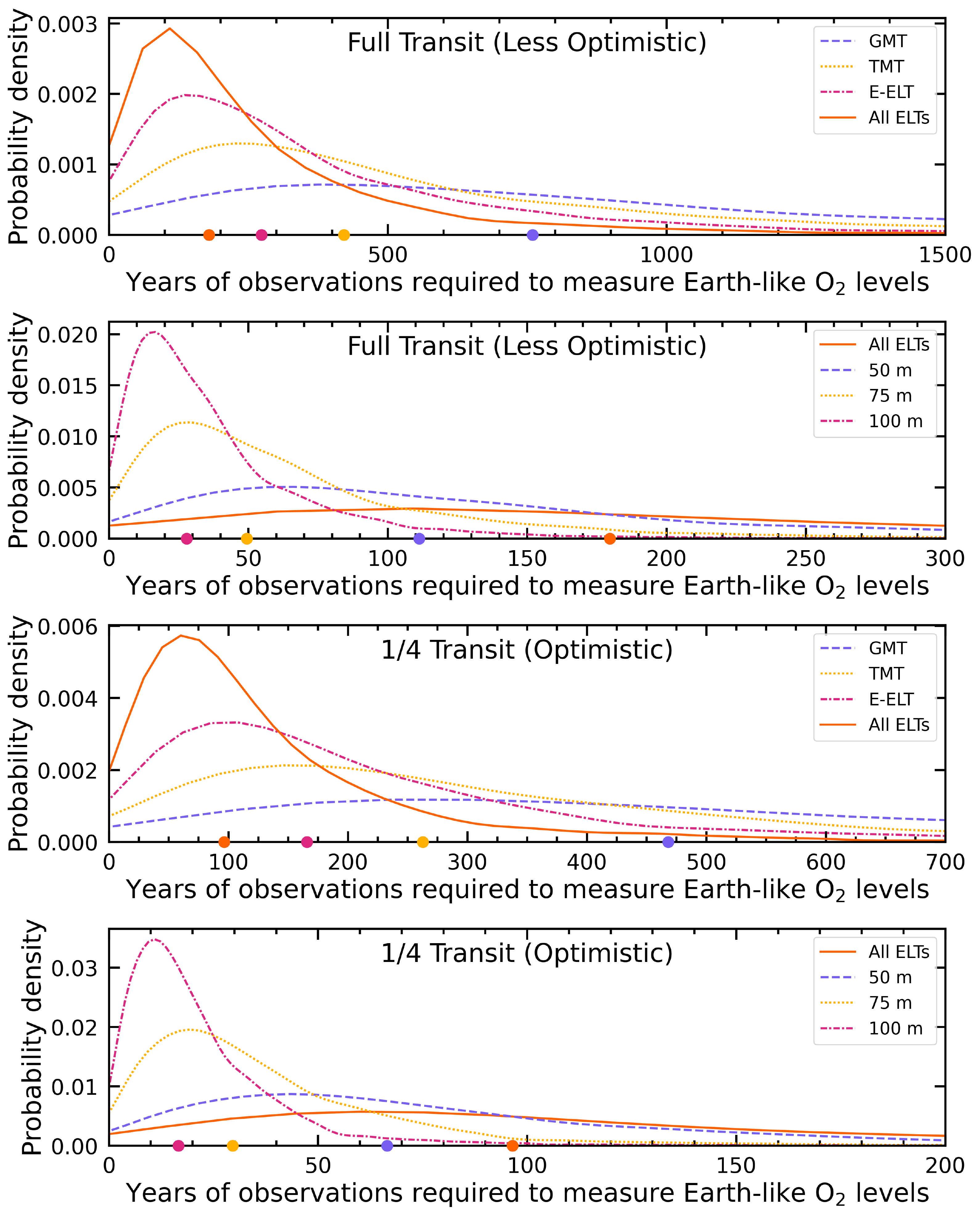}
\caption{Smoothed histograms of the number of years required to test for Earth-like \ot\ levels on hypothetical Earth-sized habitable zone planets using planned ELT telescopes and theoretical very large aperture ground-based telescopes, assuming full transits (upper two panels) and partial transits (lower two panels) are required for measurements. These calculations are based on a Monte Carlo simulation of M dwarf systems within 20\,pc, randomly sampling a fraction of stars and their hypothetical planets 10,000 times accounting for both geometric transit probability and \ee\ from \citet{Dressing2015}. The median values of the corresponding color histogram are shown as circles on the x-axis. Note the different scales along the x-axes. \label{fig:years}}
\end{figure*}

As discussed above, the interplay of a multitude of factors determines the likely planet sample size for which Earth-like \ot\ levels can be tested. Some of these factors carry significant uncertainties or are intrinsically stochastic. Therefore, we opted to use a Monte Carlo approach to simulate different realizations of hypothetical nearby transiting planet populations which we could test for Earth-like \ot\ levels. Using \texttt{Bioverse}, we randomly sampled the 20\,pc stellar population 10,000 times. In each iteration we selected a fraction of stars equal to a randomly selected \ee\ value from the distribution \ee\ $=0.16^{+0.17}_{-0.07}$ from \citet{Dressing2015} for M dwarfs, and further reduced that number based on whether or not the simulated system had a transiting habitable zone Earth-like planet. Although most of our targets are mid-M dwarfs, the \citet{Dressing2015} \ee\ measurement is the only published estimate we have for any M dwarfs and it spans a relatively broad range of values similar to those outlined in Section~\ref{sec:ee}. Each simulated ``universe'' yielded between zero and 13 stars with planets that meet our criteria to test for Earth-like \ot\ levels. For each ELT, the combined signal from multiple ELTs, and hypothetical 50, 75, and 100-meter diameter telescopes, we calculated the time span, in years, that it would take to test for Earth-like \ot\ levels on the hypothetical planets and show the resultant distributions across all simulated universes for both full and partial transits in Figure~\ref{fig:years}. These simulations suggest that on the median target in the simulation it could take nearly 100 years (partial transits) to 200 years (full transits) to test for Earth-like \ot\ levels in the best-case observing scenario by combining the signal from all ELTs. However, testing for Earth-like \ot\ levels could be possible on the median simulation target in 50 years with a telescope aperture 75 meters or larger. The results from the median simulation target are not promising for ELTs; however, the distributions in Figure~\ref{fig:years} show that some of the simulated targets could be tested for Earth-like \ot\ levels within 50 years. We summarize the fraction of simulations across all universes for which we could test for Earth-like \ot\ levels in intervals of five years up to 50 years for different ELT configurations in Tables~\ref{tab:full} and \ref{tab:quarter}.

\begin{figure*}[ht!]
\includegraphics[width=0.99\linewidth]{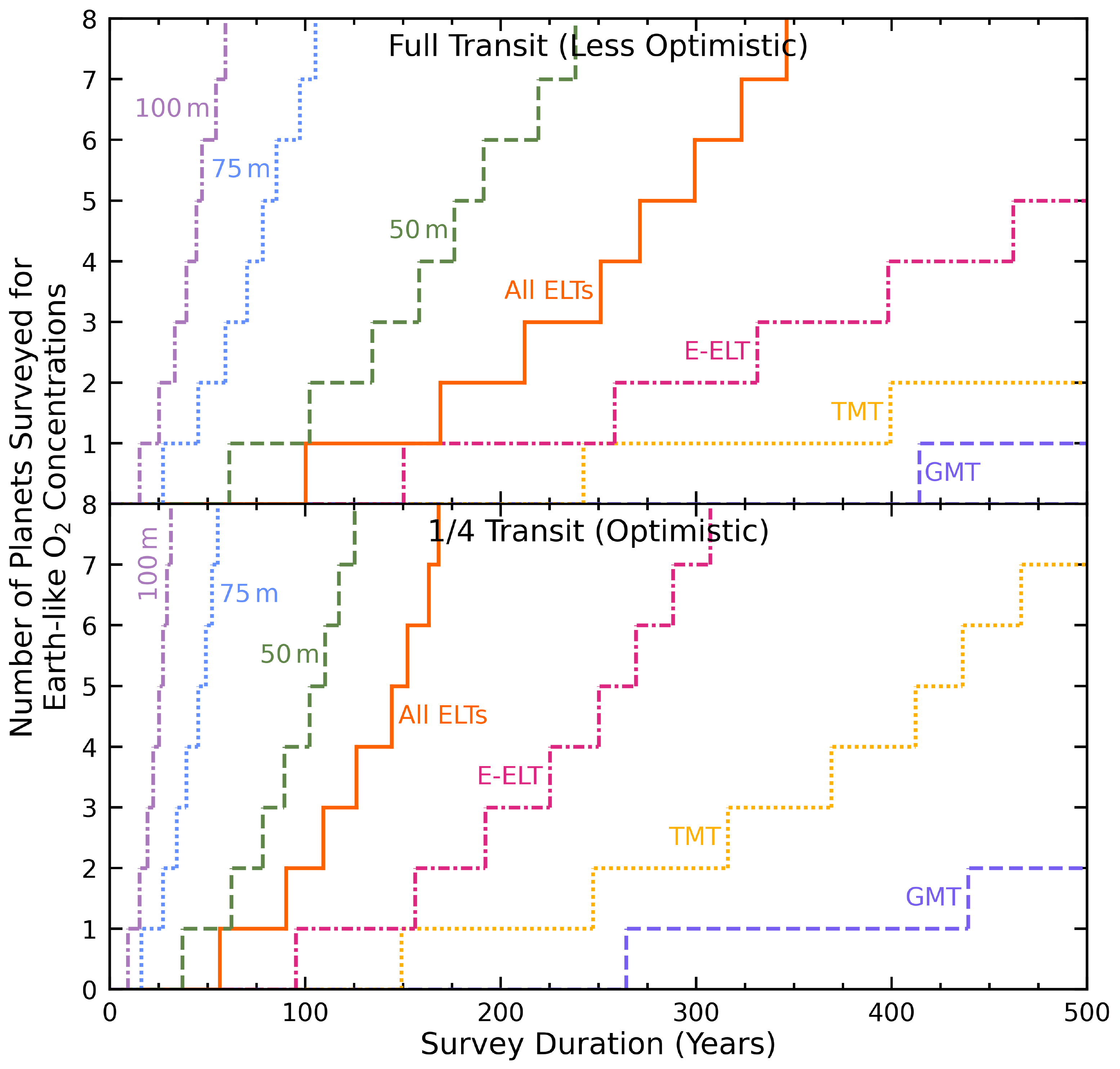}
\caption{Results from a simulated survey of hypothetical Earth-sized habitable zone planets within 20\,pc considering both \ee\ and geometric transit probability showing number of planets surveyed for Earth-like \ot\ concentrations as a function of survey duration with upcoming ELTs, combined signals from all ELTs, and hypothetical 50 to 100-meter telescopes. The upper panel shows a scenario in which full transits are required, and the lower panel shows a more optimistic scenario in which only partial transits are required. \label{fig:survey}}
\end{figure*}

From our Monte Carlo simulation, we sorted each universe by the number of years it would take to test for Earth-like \ot\ levels on the simulated planets with each ELT. We then took the median value of the first planet (smallest number of years) across all simulated universes. We repeated this for the second through thirteenth planets across universes. Most universes contain at least one planet, but progressively fewer universes contain two or more planets. Above eight planets, the median number of years to test for Earth-like \ot\ levels with some ELTs does not necessarily increase due to the significantly smaller sample of planets. We illustrate the results from these simulations in Figure~\ref{fig:survey}, showing the expected number of planets surveyed for Earth-like \ot\ concentrations as a function of survey duration. These simulations show that it could take nearly 60 years to test \ot\ levels on a single nearby Earth-sized habitable zone planet if we are able to combine signals from the upcoming ELTs in the optimistic case where we only require partial transits to be observed. If we had a 50-meter diameter or larger telescope, Earth-like \ot\ levels could be tested on a single planet in 40 years or less. Indeed, concept telescopes like the Large Fiber Array Spectroscopic Telescope \citep[LFAST,][]{Angel2022} with a proposed collecting area up to the equivalent of a $\sim$80-meter telescope could expedite \ot\ searches on one to three planets to between about 15 and 40 years, assuming it can obtain $R\approx100,000$ spectra. However, it is unclear at this time how combining measurements from hundreds of small aperture telescopes will affect the signal.

\begin{deluxetable*}{c|rrrrrrr}
\tablecaption{Fraction of Monte Carlo simulations of M dwarf systems out to 20\,pc for which Earth-like \ot\ levels could be tested on hypothetical transiting habitable zone planets within five to 50 years considering both \ee\ and geometric transit probabilities (\emph{full} transit). \label{tab:full}}
\tablehead{\colhead{Years} & \colhead{GMT} & \colhead{TMT} & \colhead{E-ELT} & \colhead{ELTs} & \colhead{50-m} & \colhead{75-m} & \colhead{100-m} } 

\startdata
5 & 0.07\% & 0.04\% & 0.22\% & 0.51\% & 0.83\% & 2.29\% & 4.47\% \\
10 & 0.15\% & 0.27\% & 0.58\% & 1.23\% & 2.00\% & 5.17\% & 11.27\% \\
15 & 0.24\% & 0.48\% & 0.97\% & 1.93\% & 3.19\% & 8.96\% & 19.42\% \\
20 & 0.37\% & 0.74\% & 1.33\% & 2.71\% & 4.47\% & 13.32\% & 27.56\% \\
25 & 0.48\% & 0.97\% & 1.72\% & 3.51\% & 6.05\% & 17.87\% & 34.95\% \\
30 & 0.60\% & 1.23\% & 2.11\% & 4.38\% & 7.66\% & 22.56\% & 41.11\% \\
35 & 0.75\% & 1.41\% & 2.48\% & 5.26\% & 9.43\% & 27.07\% & 46.94\% \\
40 & 0.86\% & 1.69\% & 2.91\% & 6.31\% & 11.27\% & 31.37\% & 51.90\% \\
45 & 0.99\% & 1.92\% & 3.36\% & 7.38\% & 13.32\% & 35.40\% & 55.96\% \\
50 & 1.18\% & 2.17\% & 3.88\% & 8.51\% & 15.58\% & 39.02\% & 59.11\% \\
\enddata
\vspace{-48pt}
\end{deluxetable*}
\vspace{-24pt}

\begin{deluxetable*}{c|rrrrrrr}
\tablecaption{Fraction of Monte Carlo simulations of M dwarf systems out to 20\,pc for which Earth-like \ot\ levels could be tested on hypothetical transiting habitable zone planets within five to 50 years considering both \ee\ and geometric transit probabilities (\emph{partial} transit). \label{tab:quarter}}
\tablehead{\colhead{Years} & \colhead{GMT} & \colhead{TMT} & \colhead{E-ELT} & \colhead{ELTs} & \colhead{50-m} & \colhead{75-m} & \colhead{100-m} } 

\startdata
5 & 0.10\% & 0.18\% & 0.47\% & 1.09\% & 1.64\% & 4.42\% & 9.79\% \\
10 & 0.27\% & 0.63\% & 1.15\% & 2.58\% & 3.86\% & 11.41\% & 24.77\% \\
15 & 0.50\% & 1.09\% & 1.93\% & 4.08\% & 6.58\% & 19.89\% & 40.63\% \\
20 & 0.74\% & 1.58\% & 2.75\% & 5.96\% & 9.79\% & 28.92\% & 53.80\% \\
25 & 0.98\% & 2.03\% & 3.60\% & 8.11\% & 13.20\% & 37.94\% & 63.93\% \\
30 & 1.20\% & 2.53\% & 4.57\% & 10.35\% & 16.90\% & 46.07\% & 70.64\% \\
35 & 1.48\% & 3.02\% & 5.54\% & 12.87\% & 20.77\% & 53.12\% & 76.12\% \\
40 & 1.75\% & 3.58\% & 6.75\% & 15.26\% & 24.77\% & 59.31\% & 80.06\% \\
45 & 2.07\% & 4.15\% & 7.77\% & 18.08\% & 28.92\% & 64.44\% & 82.92\% \\
50 & 2.31\% & 4.77\% & 9.07\% & 20.79\% & 33.18\% & 68.27\% & 85.02\% \\
\enddata
\vspace{-48pt}
\end{deluxetable*}
\vspace{-24pt}

\subsection{Case Study: Known Nearby Transiting Habitable Zone Planets}\label{sec:nhz}

Given the expected operational support lifetime of the upcoming ELTs between 30 and 50 years,\footnote{\url{https://noirlab.edu/public/media/archives/brochures/pdf/brochure023.pdf}, \url{https://elt.eso.org/about/faq/\#question\_11}, \url{https://www.tmt.org/download/Document/15/original}, \newline accessed 5 April 2023.} our simulations above suggest it is extremely unlikely we will be able to survey any planets for Earth-like \ot\ concentrations via transmission spectroscopy. However, there are two exoplanet systems within 20\,pc that host planets in their habitable zones: LHS~1140 and TRAPPIST-1.

LHS~1140 is an M4.5~V star at 15\,pc hosting at least two planets, including a $1.635\,R_{\oplus}$ super-Earth which could have surface liquid water \citep{Dittmann2017,Ment2019,Lillo-Box2020}, however, as \citet{Kimura2022} suggest at this radius the planet is likely to have a thick hydrogen-rich atmosphere that could obscure other molecules in the atmosphere. TRAPPIST-1 is an M8~V star at 12.5\,pc hosting at least seven Earth-sized planets, including four in or near the habitable zone which could have surface liquid water \citep{Gillon2016,Gillon2017}. The discovery of TRAPPIST-1 both drives the search for planets and defies current planet occurrence rate constraints for the latest-type stars (see Section~\ref{sec:ee}). The bulk composition, formation and evolutionary history, and the potential habitability of these worlds are not well constrained by current data and remain an intense area of study. While we await atmospheric constraints for the habitable zone TRAPPIST-1 planets, early results from JWST suggest the \textit{non}-habitable zone planet TRAPPIST-1~b has little or no atmosphere \citep{Greene2023}. Nevertheless, the most optimistic scenario currently conceivable is that all four TRAPPIST-1 planets in or at the edge of the habitable zone are habitable and are broadly Earth-like worlds, where an \ot\ search is compelling. Therefore, we investigated this possibility.  

Tables~\ref{tab:fullyr} and \ref{tab:partyr} list the results from our simulated habitable zone planets, however, for TRAPPIST-1 and LHS~1140, we also ran our calculations on the known and already detected habitable zone planets in those systems. It would take $\sim$60 years to test for Earth-like levels of \ot\ on LHS~1140~b, assuming we observed all partial transits of the planet from each ELT and combined the signal, 130 years if full transits are required. The TRAPPIST-1 planets are more promising and could take between $\sim$16 and $\sim$55 years to test for Earth-like levels of \ot\ on planets d through g, respectively, assuming partial transits and combining signals from every observable transit from all ELTs. If full transits are required, this time scale would increase to $\sim$25 to $\sim$85 years for planets d through g, respectively, from combined measurements.

\begin{figure*}[ht!]
\includegraphics[width=0.99\linewidth]{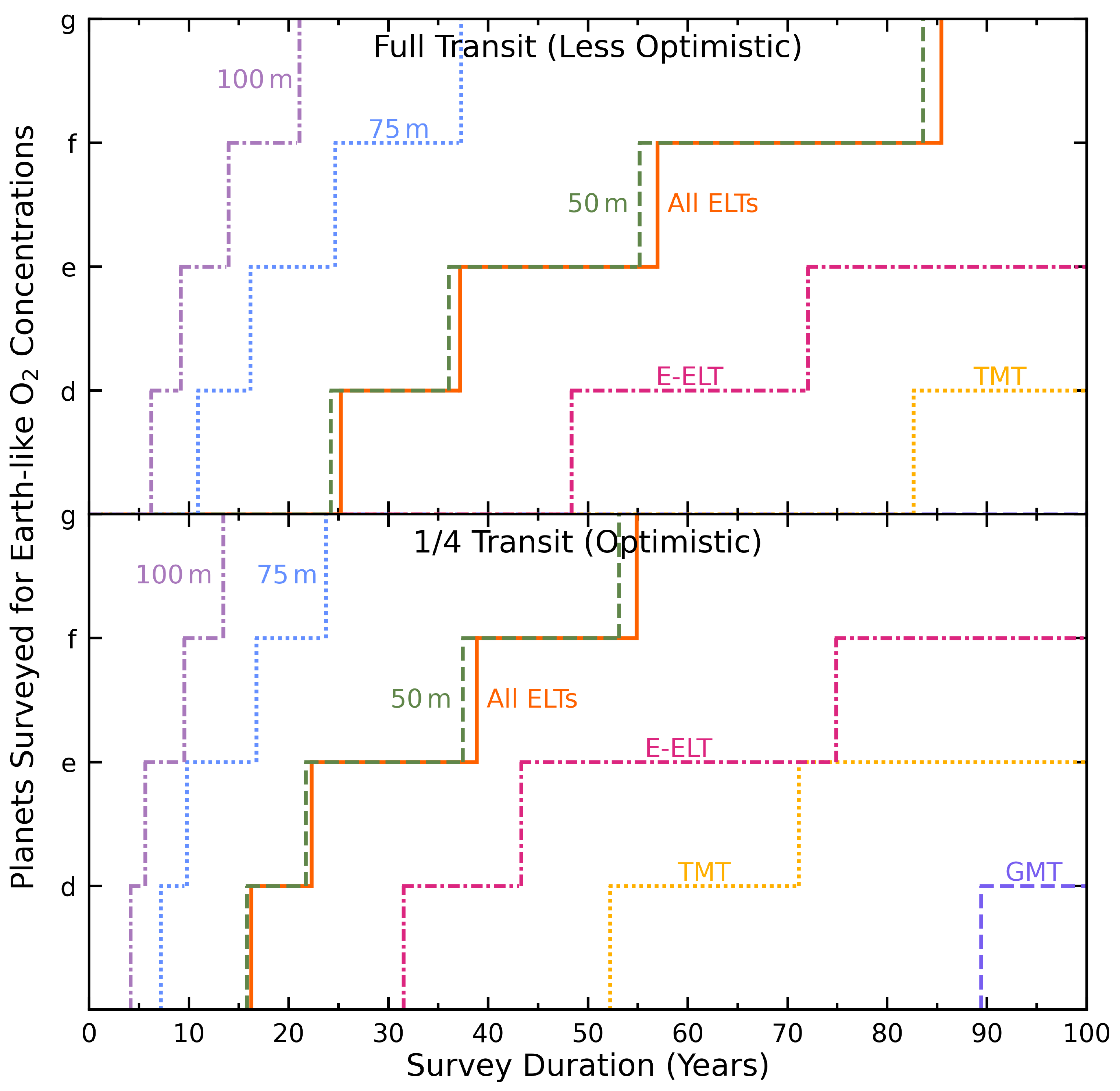}
\caption{A simulation of expected number of years to test for Earth-like levels of \ot\ on the TRAPPIST-1 habitable zone planets. The upper panel shows the expected time to measure \ot\ for different telescopes if full transits are required, and the lower panel shows the same results if only partial transits are required. \label{fig:trappist}}
\end{figure*}

As an example of an optimistic scenario of measuring Earth-like levels of \ot\ on nearby habitable zone Earth-sized planets, we summarize the results for the TRAPPIST-1 system planets in Figure~\ref{fig:trappist} for the different observing modes used in Figure~\ref{fig:years}. Observations with JWST can reveal whether or not the TRAPPIST-1 planets have atmospheres and if it would even be worth pursuing observations of \ot\ on these planets and similar planets yet to be found orbiting late-type stars.

\subsection{Space vs.\ Ground-based Telescope Searches}

Ultimately, the yields predicted by our simulations are limited primarily by the occurrence rates of nearby transiting Earth-sized habitable zone planets and the photon noise achievable with the largest currently conceivable telescopes. From a superficial assessment, it may seem that our results would also indicate that \textit{space-based} detection of \ot\ in transiting planets would be similarly challenging. It is very important to stress that this is \textit{not} the case. Our results are not applicable to space-based observations, due to two fundamental differences: First, because space-based telescopes observe from above the terrestrial atmosphere, there is no need to distinguish telluric spectral lines from exoplanet spectral lines. Therefore, there is no need to rely on systemic velocity differences between the lines and, consequently, there is no need for very high-resolution spectroscopy. From space, low- and medium-resolution spectroscopy have important advantages, as they can provide higher information content (considering signal-to-noise ratios) and will require fewer photons for a detection. Furthermore, as there are no constraints on the relative systemic velocity of the target planet and Earth, a greater number of targets will be accessible.

Second, space-based observations can provide orders-of-magnitude more stable--thermally and mechanically--environments than ground-based observatories. This important difference (exploited by \textit{Kepler}, \textit{TESS}, \textit{HST} and \textit{JWST}) makes time-resolved high-precision intensity measurements easier from space than from the ground.

Given the two considerations above, space-based transit observations will be much more productive if utilizing low-resolution transit spectroscopy. However, a telescope like \textit{JWST} is not well suited for \ot\ detection in exoplanet atmospheres. \citet{Fauchez2020} determined that a 6.4~$\mathrm{\mu}$m \ot--N$_2$ collision-induced absorption feature could be detectable with JWST over five years (the original expected JWST mission duration) on a TRAPPIST-1~e analog within 5\,pc of the Sun in 85 transits. At the distance of TRAPPIST-1, this would take nearly 10 times longer, $\sim$800 transits. If \textit{JWST} was looking at the \ot\ A band or 1.27~$\mathrm{\mu}$m band, it would take $\sim$2,000 transits for a 5\,pc TRAPPIST-1 analog.

The Astro2020 Decadal Survey recommended an at least 6-meter diameter IR/O/UV space telescope, capable of observing $\sim$100 nearby exoplanet systems, and optimized for observing habitable exoplanets via direct imaging \citep{NationalAcademiesofSciences2021}. The ultimate goal of the high-contrast imaging mission--currently referred to as the Habitable Worlds Observatory--may be to probe for Earth-like \ot\ levels in about 25 habitable planets. However, a space telescope with such an aperture size and mission duration similar to JWST is not likely to be capable of efficient \ot\ surveys in \textit{transiting} planets explored in our study.

More ambitious multiple-aperture space telescopes might be required for robust \ot\ detection, confirmation, and synergy with ELTs in the coming decades. The Large Interferometer For Exoplanets (LIFE) mission concept proposes an array of four 2--3-meter class apertures in space working at mid-infrared wavelengths to measure thermal emission of exoplanets \citep{Quanz2022,Kammerer2022}. The LIFE telescope would not be able to constrain \ot\ concentrations, but might be able measure Earth-like levels of O$_3$ and possibly CH$_4$ on an Earth analog around a G2V star at 10\,pc \citep{Konrad2022}.

The new Space 2.0 technology--propelled by new, reusable launch vehicles and intensifying international competition--may open new pathways. The Nautilus Space Observatory concept \citep{Apai2019} is a space telescope constellation consisting of thirty-five 8.5-meter diameter telescopes optimized for transiting planet spectroscopy. The telescopes use a novel optical element, multi-order diffractive engineered material lenses \citep[][]{Milster2020,Milster2021}, that provide large telescope diameters that are cost-effective, replicable, and have high image quality over a broad wavelength range.  These telescopes could be mass produced and launched in groups, providing a resilient, scalable constellation. These telescopes would operate independently, but the signal could be digitally co-added to achieve the light-collecting power of a 50-meter space telescope \citep{Apai2019}. With a full 35-element array, \citet{Apai2019} predicted that biosignatures and key atmospheric absorbers (O$_3$, H$_2$O, and \ot) at Earth-like concentrations can be searched for in the atmospheres of $\sim$1,000 exo-Earth analogs orbiting M dwarfs, K dwarfs, or G dwarfs. This sample would be probed within the mission's 10-year lifetime. The \citet[][]{Apai2019} study used a similar analysis to the current paper. It was optimistic in the sense that it assumed cloud-free atmospheres and a very simplistic instrument model. Scaling from that study, \citet{Apai2019b} proposed that a single 8.5-meter diameter Nautilus unit (Probe-class mission) could detect evidence for \ot\ on an Earth-like planet orbiting a 15\,pc M dwarf in 30 transits. In follow-up studies \citep[][]{Bixel2020,Bixel2021}, the previous analysis was extended to include realistic cloud cover \citep[derived from 3D general circulation models,][]{Komacek2019}, considering a reduced system-level throughput, but without the use of detailed instrument models. Including clouds resulted in a reduced yield of $\sim$200 exo-Earths probed for Earth-like concentrations of \ot\ within a four year mission lifetime. (For cloud-free atmospheres, about 800 exo-Earths can be probed within a ten year mission lifetime.) Future studies of this concept should include more detailed instrument models and account for the range of atmospheric compositions/cloud conditions in the target sample.

Given the above, it is possible to project a potential future scenario under the assumption that all the above facilities are built on optimistic, but not unrealistic timelines. We will assume that ELTs and their relevant instruments become fully operational and begin their surveys by 2030, the Habitable Worlds Observatory and LIFE are launched in 2040, LFAST begins its survey in 2045 and Nautilus Space Observatory is launched in the same year. With such a timeline, the Habitable Worlds Observatory would probe up to 25 (directly imaged) planets for \ot\ by 2050. By that time, the joint operations of the ELTs may have provided another 1--3 targets to the sample, \textit{if} the TRAPPIST-1 planets are indeed habitable. By 2050, LIFE could follow up on some of the HWO-observed planets. The LFAST project could survey another 3--5 targets by 2055. By that year, the Nautilus Space Observatory could boost the number of planets surveyed from $\sim$30 to $\sim$400, and--by 2060--reach $\sim$1,000 planets probed, allowing large-scale statistical studies of biosignatures. 

If the ELTs are to contribute to the above timeline, planning, coordination, and collaboration between the telescopes and among institutional partners would be crucial. Since decade-long monitoring campaigns will be necessary for \ot\ detection via transmission spectroscopy on a small number of targets, a special request to allocate uninterruptible observing time for these targets would likely need to be made. Given that a single habitable-zone Earth-like planet orbiting an M dwarf will have fewer than about 10 observable transits per year, and the transit duration is on the order of 1--3 hours, this request does not seem unreasonable for a small number of targets, especially due to the importance of the discovery of a potential biosignature on an Earth-analog.

A final note that could drastically alter the scope of ground-based observations, and one key factor we did not fully consider in our analysis, is the growing number of Earth-orbiting satellite constellations. A crucial component of many large ground-based telescopes is the multi-conjugate adaptive optics system that mitigates atmospheric turbulence. These systems employ deformable mirrors and powerful lasers to create multiple artificial guide stars to measure and correct for atmospheric turbulence and for many telescope/instrument misalignments. A few days prior to observations, a target list is sent to the Department of Defense Laser Clearinghouse which determines when the laser systems must be turned off during observations. However, with the anticipated launch of tens to hundreds of thousands of new satellites in the next decade, we expect satellites to drastically affect observations if laser systems must be shuttered or shut off frequently. Observations could also be affected if laser operation is limited to certain altitudes above the horizon or times of night due to air traffic. The ELTs currently plan to use software to minimize laser shuttering time due to passing satellites and airplanes \citep{Hainaut2022}.

\section{Summary and Conclusions}\label{sec:summary}

By creating a large and homogeneous stellar catalog of nearby stars and integrating it into the \texttt{Bioverse} mission analysis framework, we enable studies of systems most amenable to follow-up efforts by upcoming telescopes and space missions. The following list summarizes the main conclusions of this work:

\begin{itemize}
    \item We computed basic stellar parameters (\teff, luminosity, radius, and mass) for 286,391 main sequence stars out to 120\,pc from the Gaia Catalogue of Nearby stars in a uniform manner using \gaia\ DR3 astrometry and photometry (Table~\ref{tab:starpars} and Figure~\ref{fig:gcnscmd}). We also computed systemic velocities for the $\sim$40\% of these targets without spectroscopic systemic velocities from \gaia\ using a novel random forest regression approach, yielding an uncertainty of 17.51~km~s$^{-1}$ for these measurements.
    
    \item We compared our photometric \teff, radius, and mass measurements to those from both spectroscopic observations and planet host measurements from the literature and we found typical consistency within $\sim$5\% for all measurements (Figure~\ref{fig:plcomp}).
    
    \item We used \texttt{Bioverse} to estimate the number of transiting Earth-sized habitable zone planets within 20\,pc, considering different scenarios spanning a conservative range of literature values for \ee. Only in the most optimistic cases for \ee\ (e.g., \ee$_{,FGKM}$=35\%) do we expect there to be one or more of these planets to be discovered within 10\,pc, most likely orbiting an M dwarf (Figure~\ref{fig:eta}).
    
    \item We updated the \texttt{Bioverse} framework to incorporate the 120\,pc stellar catalog and add additional functionality. We then used \texttt{Bioverse} to simulate hypothetical transiting Earth-sized planets ($0.8S^{0.25} < R_p < 1.3\,R_{\oplus}$) in the habitable zones of main sequence stars within 20\,pc and computed how many transits of these hypothetical planets would be observable from each of the three upcoming ELTs accounting for relative system velocity in order to compute \ot\ \textit{A} and IR band observability and telescope site weather (Tables~\ref{tab:fulltr} and \ref{tab:parttr}). Our Monte Carlo simulations yielded more than one observable transit per year, on average, from ground-based ELTs for stars within 20\,pc later than spectral type M2.5~V, if they hosted a habitable zone planet. However, we expect 16 or fewer of these transiting habitable zone Earth analogs to be discovered (Figures~\ref{fig:o2obs} and \ref{fig:o2obsq}).
    
    \item Taking into consideration \ee\ estimates of $0.16^{+0.17}_{-0.07}$ from \citet{Dressing2015}, geometric transit probability, target brightness, and models of sensitivity of first-generation high-resolution spectrographs on upcoming ELTs, we ran a Monte Carlo simulation to generate populations of hypothetical transiting Earth-sized habitable zone planets around M dwarfs within 20\,pc in order to determine a scenario for measuring Earth-like \ot\ levels. Tables~\ref{tab:full} and \ref{tab:quarter} summarize the fraction of simulations for which Earth-like \ot\ levels could be tested in hypothetical Earth analogs. Based on the median simulated planet, it is unlikely Earth-like levels of \ot\ could be probed with upcoming ELTs within 50 years, even if signals from multiple ELTs were combined (Figure~\ref{fig:survey}). However, if $R=500,000$ spectrographs were used, this would expedite the time to probe for Earth-like \ot\ levels by a factor of two (Figure~\ref{fig:survey5k}).
    
    \item If there are clear atmospheres on TRAPPIST-1 d--g, Earth-like \ot\ concentrations could be probed in a very optimistic case by combining signals from each observable partial transit from all three upcoming ELTs within 16 to 55 years (Figure~\ref{fig:trappist}).
\end{itemize}

Probing for Earth-like levels of \ot\ on transiting nearby habitable zone Earth analogs from ground-based ELTs via transmission spectroscopy will take decades with currently planned instrumentation, even in the most optimistic case. This makes a survey for \ot\ from ground-based telescopes unlikely in the near-future unless very compelling targets are discovered. Transmission spectroscopy with ELTs will still be a fruitful endeavor, enabling exciting and crucial atmospheric studies of hundreds of exoplanets. The search for biosignatures on habitable zone planets with ground-based ELTs will likely continue, however, primarily via direct imaging rather than transmission spectroscopy.

\section{acknowledgments}

We acknowledge discussions with and contributions by Alex Bixel and Sam Myers to an early exploratory analysis of questions similar to those explored here. KKH-U acknowledges Megan Mansfield and Martin Schlecker for general discussions about transmission spectroscopy, code, and planet formation.

This material is based upon work supported by the National Aeronautics and Space Administration under Agreement No. 80NSSC21K0593 for the program ``Alien Earths''. The results reported herein benefited from collaborations and/or information exchange within NASA's Nexus for Exoplanet System Science (NExSS) research coordination network sponsored by NASA’s Science Mission Directorate. IP and GJB acknowledge support from the NASA Astrophysics Data Analysis Program under grant No. 80NSSC20K0446.

This research has made use of the NASA Exoplanet Archive, which is operated by the California Institute of Technology, under contract with the National Aeronautics and Space Administration under the Exoplanet Exploration Program.

This work has made use of data from the European Space Agency (ESA) mission {\it Gaia} (\url{https://www.cosmos.esa.int/gaia}), processed by the {\it Gaia} Data Processing and Analysis Consortium (DPAC, \url{https://www.cosmos.esa.int/web/gaia/dpac/consortium}). Funding for the DPAC has been provided by national institutions, in particular the institutions participating in the {\it Gaia} Multilateral Agreement.

This research has made use of the SIMBAD database,
operated at CDS, Strasbourg, France.

Funding for the Sloan Digital Sky Survey IV has been provided by the Alfred P. Sloan Foundation, the U.S. Department of Energy Office of Science, and the Participating Institutions. 

SDSS-IV acknowledges support and resources from the Center for High Performance Computing at the University of Utah. The SDSS website is www.sdss.org.

SDSS-IV is managed by the Astrophysical Research Consortium for the Participating Institutions of the SDSS Collaboration including the Brazilian Participation Group, the Carnegie Institution for Science, Carnegie Mellon University, Center for Astrophysics | Harvard \& Smithsonian, the Chilean Participation Group, the French Participation Group, Instituto de Astrof\'isica de Canarias, The Johns Hopkins University, Kavli Institute for the Physics and Mathematics of the Universe (IPMU) / University of Tokyo, the Korean Participation Group, Lawrence Berkeley National Laboratory, Leibniz Institut f\"ur Astrophysik Potsdam (AIP),  Max-Planck-Institut f\"ur Astronomie (MPIA Heidelberg), Max-Planck-Institut f\"ur Astrophysik (MPA Garching), Max-Planck-Institut f\"ur Extraterrestrische Physik (MPE), National Astronomical Observatories of China, New Mexico State University, New York University, University of Notre Dame, Observat\'ario Nacional / MCTI, The Ohio State University, Pennsylvania State University, Shanghai Astronomical Observatory, United Kingdom Participation Group, Universidad Nacional Aut\'onoma de M\'exico, University of Arizona, University of Colorado Boulder, University of Oxford, University of Portsmouth, University of Utah, University of Virginia, University of Washington, University of Wisconsin, Vanderbilt University, and Yale University.

This work made use of the Third Data Release of the GALAH Survey \citep{Buder2021}. The GALAH Survey is based on data acquired through the Australian Astronomical Observatory, under programs: A/2013B/13 (The GALAH pilot survey); A/2014A/25, A/2015A/19, A2017A/18 (The GALAH survey phase 1); A2018A/18 (Open clusters with HERMES); A2019A/1 (Hierarchical star formation in Ori OB1); A2019A/15 (The GALAH survey phase 2); A/2015B/19, A/2016A/22, A/2016B/10, A/2017B/16, A/2018B/15 (The HERMES-TESS program); and A/2015A/3, A/2015B/1, A/2015B/19, A/2016A/22, A/2016B/12, A/2017A/14 (The HERMES K2-follow-up program). We acknowledge the traditional owners of the land on which the AAT stands, the Gamilaraay people, and pay our respects to elders past and present. This paper includes data that has been provided by AAO Data Central (datacentral.org.au).

Guoshoujing Telescope (the Large Sky Area Multi-Object Fiber Spectroscopic Telescope LAMOST) is a National Major Scientific Project built by the Chinese Academy of Sciences. Funding for the project has been provided by the National Development and Reform Commission. LAMOST is operated and managed by the National Astronomical Observatories, Chinese Academy of Sciences.

\vspace{5mm}
\begin{large}\textit{Author contributions:}\end{large}
KKH-U and DA developed the research project. DA initiated and funded the project and proposed the initial version of the methodology. KKH-U built the target catalog, performed the calculations and analysis, and drafted the manuscript. DA made major contributions to the manuscript. GJB and IP provided guidance and input regarding \ee\, and GJB assisted with the luminosity calculations. ML-M provided the instrument models to determine how many transits are necessary to probe for Earth-like levels of \ot. All authors contributed edits and suggestions to improve the manuscript.

\vspace{5mm}
\facilities{Exoplanet Archive, Gaia, Sloan (APOGEE), UKST (GALAH), LAMOST}

\vspace{5mm}
\software{\texttt{astroplan} \citep{astroplan2018}, \texttt{astropy} \citep{astropy:2013, astropy:2018, AstropyCollaboration2022}, \texttt{Bioverse} \citep{Bixel2021}, \texttt{matplotlib} \citep{Hunter:2007}, \texttt{numpy} \citep{harris2020array}, \texttt{pandas} \citep{mckinney-proc-scipy-2010,reback2020pandas}, \texttt{scikit-learn} \citep{Pedregosa2011}}

\appendix

\section{Simulations for an \texorpdfstring{$R=500,000$}{R=500,000} spectrograph}\label{app:500k}

\begin{figure*}[ht!]
\includegraphics[width=0.98\linewidth]{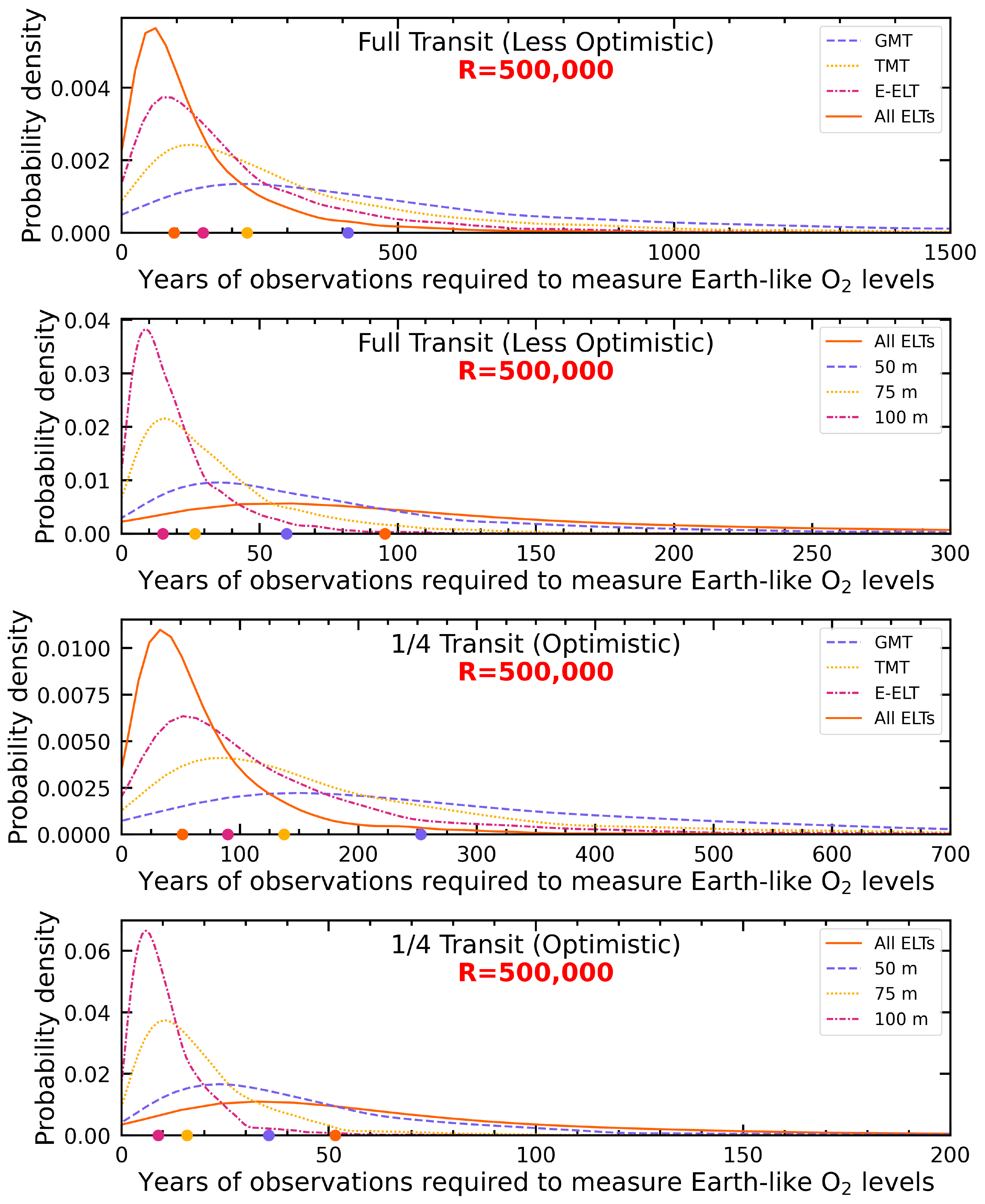}
\caption{Smoothed histograms of the number of years required to test for Earth-like \ot\ levels on hypothetical Earth-sized habitable zone planets using planned ELT telescopes and theoretical very large aperture ground-based telescopes with $R=500,000$ spectrographs, assuming full transits (upper two panels) and partial transits (lower two panels) are required for measurements. These calculations are based on a Monte Carlo simulation of M dwarf systems within 20\,pc, randomly sampling a fraction of stars and their hypothetical planets 10,000 times accounting for both geometric transit probability and \ee\ from \citet{Dressing2015}. The median values of the corresponding color histogram are shown as circles on the x-axis. Note the different scales along the x-axes. \label{fig:years5k}}
\end{figure*}

\begin{figure*}[ht!]
\includegraphics[width=0.99\linewidth]{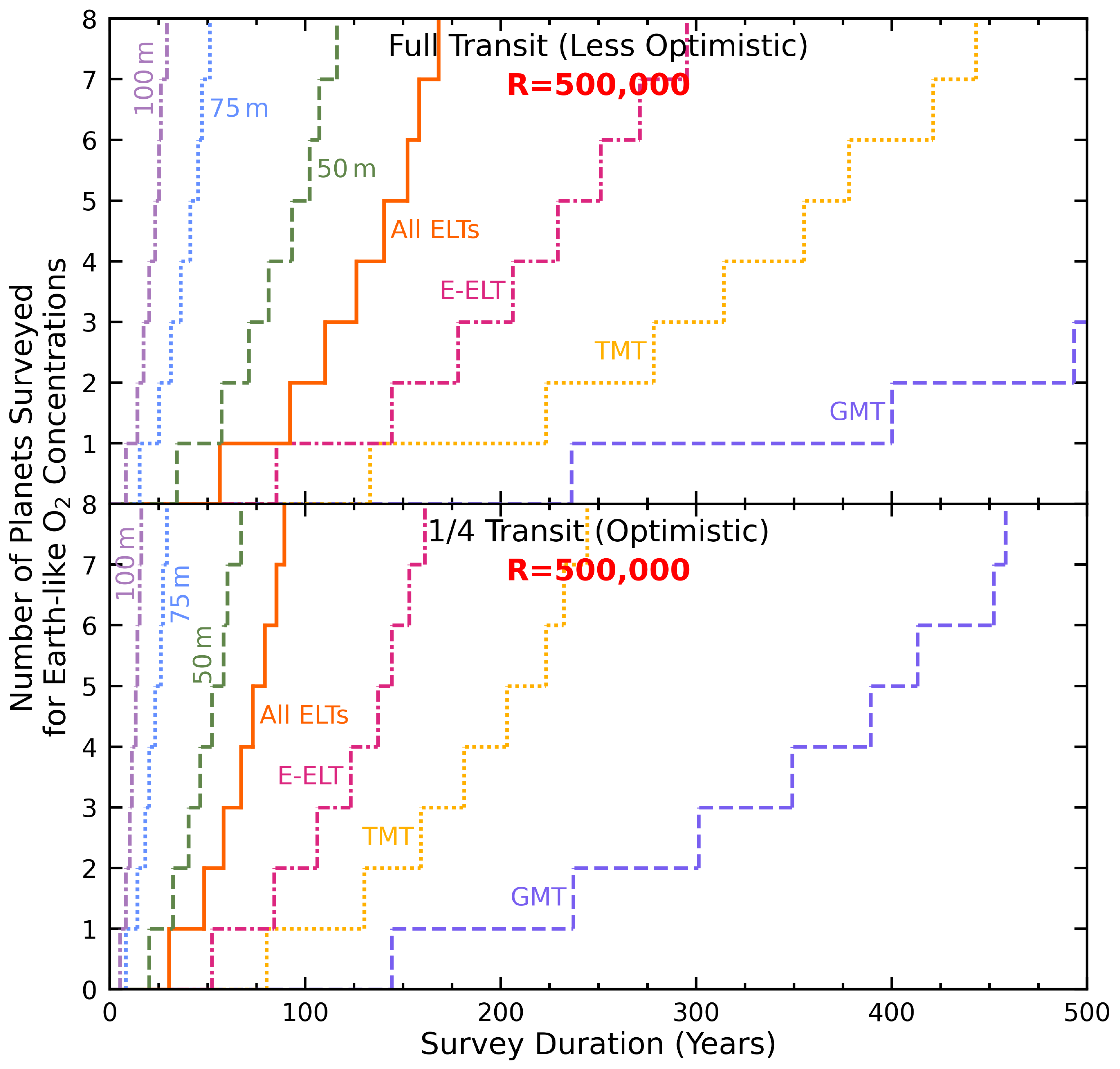}
\caption{Results from a simulated survey of hypothetical Earth-sized habitable zone planets within 20\,pc considering both \ee\ and geometric transit probability showing number of planets surveyed for Earth-like \ot\ concentrations as a function of survey duration with upcoming ELTs, combined signals from all ELTs, and hypothetical 50 to 100-meter telescopes with $R=500,000$ spectrographs. The upper panel shows a scenario in which full transits are required, and the lower panel shows a more optimistic scenario in which only partial transits are required. \label{fig:survey5k}}
\end{figure*}

The calculations in this paper are based on the nominal resolution of $R=100,000$ of currently planned first-generation high resolution spectrographs for the ELTs. However, as \citet{Lopez-Morales2019} noted, at spectral resolution higher than $R=300,000$ the average \ot\ line depth is more than double and line blending is less severe. As such, in this Appendix we ran our above analysis for a hypothetical $R=500,000$ spectrograph based on the highest resolution models from \citet{Lopez-Morales2019}. Figures~\ref{fig:years5k} and \ref{fig:survey5k} are the same as Figures~\ref{fig:years} and \ref{fig:survey}, but for an $R=500,000$ spectrograph. The $R=500,000$ plots indicates that such an instrument would reduce the time to survey for \ot\ by a factor of $\sim$2.

\bibliography{main}{}
\bibliographystyle{aasjournal}

\end{document}